\documentclass[
aps, %
prb,
graphicx,
superscriptaddress, %
amsmath, amssymb,
twocolumn, %
reprint, %
floatfix
]{revtex4-2}
\usepackage[ascii]{inputenc}
\usepackage[T1]{fontenc}
\usepackage[USenglish]{babel}
\usepackage{dcolumn}%
\usepackage{multirow}
\usepackage[dvipsnames]{xcolor}
\usepackage{hyperref}

\usepackage{graphicx}
\usepackage[caption=false]{subfig}
\usepackage[ignoreunlbld,norefs,nocites]{refcheck}

\renewcommand{\vec}[1]{\mathbf{#1}} %
\newcommand{\intd}[1]{\mathrm{d}\vec{#1}} %
\newcommand{\fdv}[2]{\frac{\delta #1}{\delta #2 }}  %

\newcommand{\tarden}{\rho_\mathrm{t}}
\newcommand{\qmcden}{\rho_\mathrm{QMC}}
\newcommand{\qmcspinden}{\rho^\sigma_\mathrm{QMC}}
\newcommand{\vext}{v_\mathrm{ext}}
\newcommand{\vhar}{v_H}
\newcommand{\vxcs}{v_{xc}}
\newcommand{\vks}{v_s}
\newcommand{\exc}{E_{xc}}

\newcommand{\qmcdeng}[1]{\rho_{\vec{G},#1}^\mathrm{QMC}}
\newcolumntype{d}[1]{D{.}{.}{#1}} %

\begin{document}

\title{From Densities to Potentials: Benchmarking Local Exchange-Correlation Approximations}

\author{Visagan Ravindran}
\affiliation{Department of Physics, Durham University, South Road, Durham, DH1 3LE, United Kingdom}
\author{Clio Johnson}
\author{Neil D.\ Drummond}
\affiliation{Department of Physics, Lancaster University, Lancaster LA1 4YB, United Kingdom}
\author{Stewart J.\ Clark}
\author{Nikitas I.\ Gidopoulos}
\affiliation{Department of Physics, Durham University, South Road, Durham, DH1 3LE, United Kingdom}
\email[Corresponding author: ]{nikitas.gidopoulos@durham.ac.uk}

\date{30th October 2025}

\begin{abstract}
    Using the Kohn-Sham (KS) inversion method of Hollins \textit{et al.}\ [J.\ Phys.:\ Condens.\ Matter \textbf{29}, 04LT01 (2017)], we invert densities from variational and diffusion quantum Monte Carlo (QMC) calculations to obtain benchmark QMC-KS potentials for a range of insulators and semiconductors, which we then compare to the KS potentials of popular density functional approximations (DFAs).

    Our results show that different DFAs yield similar electron densities, despite differences in their KS potentials, which originate primarily
    from the exchange and correlation contribution.
    We also find that the KS gap from the QMC density is typically larger than the KS gaps of most DFAs, with the exception of Hartree-Fock.
    Finally, the KS gap is sensitive to the inclusion of semicore states in the pseudopotentials, such that comparison with experiment should be done with caution.
\end{abstract}

\maketitle

\section{Introduction}
Kohn-Sham (KS) density functional theory (DFT) \cite{Kohn-Sham:1965,Hohenberg_Kohn:1964} is ubiquitous in \textit{ab initio} electronic structure calculations within condensed matter physics, chemistry, and materials science. In KS theory, an interacting system of electrons is mapped to an auxiliary noninteracting system of particles moving in a mean-field effective potential $\vks(\vec{r})$ that is constructed such that the ground state densities of the interacting and noninteracting systems are the same.
However, the exact exchange-correlation (XC) contribution $\exc[\rho]$ to the KS energy functional is unknown,
and by extension the exact XC contribution
\begin{equation}
    \vxcs[\rho](\vec{r}) = \fdv{\exc[\rho]}{\rho(\vec{r})}
\end{equation}
to the KS potential $\vks(\vec{r})$ is also unknown, necessitating approximations for both $\exc[\rho]$ and $\vxcs(\vec{r})$ in practice.
Numerous density functional approximations (DFAs) \cite{B88,PBE,PBESOL,TPSS,SCAN,RSCAN,R2SCAN,B3LYP,HSE,PBE0}
have been proposed over the decades but the \textit{direct} assessment of the accuracy of $\exc[\rho]$ and $\vks(\vec{r})$ is a formidable challenge, not least because neither are experimentally observable quantities.
Density inversion \cite{den_inv_Leeuwen_Baerends:1994,den_inv_Gorling:1992,den_inv_Zhao_Morrison_Parr:1994,den_inv_Tozer:1995,den_inv_Savin_Umrigar_Gonze:1998,den_inv_Waroquier:2003,den_inv_Stott:2004,den_inv_Staroverov:2012,den_inv_Wasserman:2018,den_inv_Wasserman:2021,den_inv_Kumar:2019,den_inv_Gorling:2022,den_inv_Aouina_Reining_QMC:2023,den_inv_Aouina_QMC:2024,den_inv_Kuemmel_QMC_mols:2025}
offers a possible methodology to assess $\vks(\vec{r})$ and in particular $\vxcs(\vec{r})$ by tackling the inverse KS problem, where for a given target density $\tarden(\vec{r})$ one finds the KS potential $\vks(\vec{r})$ such that the ground-state density of the KS auxiliary system is equal to $\tarden(\vec{r})$.
Consequently, the inversion of numerically accurate densities enables one to gain insight into the behavior of the exact $\vks(\vec{r})$.

Quantum Monte Carlo (QMC) \cite{Foulkes:01} methods are a well-established family of many-body techniques, which have played a historically significant role in the development of DFAs, notably in facilitating parameterizations \cite{LDA_PW,LDA_PZ} for the local density approximation (LDA) through QMC
simulations of the homogeneous electron gas (HEG) \cite{Ceperley:80}.
Among such methods are the variational and diffusion Monte Carlo (VMC and DMC) methods.
In the VMC method, electron configurations are sampled from the square modulus of a trial wave function using the Metropolis algorithm, with estimators of observables of interest being averaged over those configurations.  The trial wave function contains free parameters that are optimized by minimizing either the energy expectation value or the variance of the energy.  In the DMC method  \cite{Ceperley:80} a population of electron configurations (also known as walkers) is evolved according to the Schr\"{o}dinger equation in imaginary time to project out the ground-state component of the trial wave function. Fermionic antisymmetry is maintained by fixing the complex phase of the DMC wave function at that of the VMC trial wave function.

In this work, we use the VMC and DMC methods to calculate
the ground-state electronic densities of various semiconductors and insulators.
We then invert these densities to find the KS potentials
$\vks(\vec{r})$ that give rise to them in the ground state.
Using the QMC densities and QMC-derived KS potentials as a benchmark,
we assess the quality of various DFAs using a range of metrics, including KS band gaps, integrated absolute density differences, and integrated potential differences weighted by density differences.
Following Burke and coworkers \cite{Burke_DE_FE:2013,Burke_DE_FE:2017,Burke_DE_FE:2018,Burke_DE_FE:2019}, we also analyze the total energy error of each DFA by examining the density-driven and functional error contributions.

The rest of this article is structured as follows.
In Sec.\ \ref{section:qmc_details} we provide details of our QMC calculations, in particular the calculation of densities, while Sec.\ \ref{section:inversion_algorithm} outlines our algorithm
to invert densities and obtain the KS potential.
We describe the errors and uncertainties in our charge densities in Sec.\ \ref{section:errors}.  We compare our exact Kohn-Sham potential with other local potentials in Sec.\ \ref{section:exact_KS}, while the KS band gaps for each DFA along with the XC derivative discontinuity $\Delta_{xc}$ are described in Sec.\ \ref{section:discontinuity}. Finally, we draw our conclusions in Sec.\ \ref{section:conclusions}.
We use Hartree atomic units (a.u.), in which the reduced Planck's constant $\hbar$, the electron mass $m_\text{e}$, the magnitude of the electronic charge $e$, and $4\pi\varepsilon_0$ are 1 a.u, where $\varepsilon_0$ is the permittivity of free space.

\section{Theory and computational details} %

\subsection{Obtaining QMC charge densities}\label{section:qmc_details}

\subsubsection{QMC calculations}

The VMC and DMC calculations reported in this work were performed using the \textsc{casino} program \cite{Needs:20}, and the DFT calculations were carried out using the \textsc{castep} plane-wave-basis-set code \cite{Clark:05}.
For the materials studied we used experimental lattice parameters taken from Ref.\ \onlinecite{Madelung:1996} unless otherwise stated (see Table\ \ref{table:supercell_gaps}).
Trail-Needs (TN) Dirac-Fock pseudopotentials \cite{Trail:05} were used to represent atomic cores.
The $s$ channel was chosen to be the local channel of the pseudopotential in each case to avoid possible issues with ghost states \cite{Drummond:16} that can arise due to the Kleinman-Bylander \cite{Kleinman_Bylander:1982} representation of the pseudopotentials in plane-wave DFT calculations. %

Our QMC trial wave functions were of Slater-Jastrow (SJ) form,
\begin{equation}
    \Psi_\text{T}\left(\vec{R}\right) = S_\uparrow\left(\vec{R}\right) S_\downarrow\left(\vec{R}\right)\exp\left(J\left(\vec{R}\right)\right),
\end{equation}
where $\vec{R}$ is the $3N$-dimensional electron configuration vector and $S_{\uparrow/\downarrow}\left(\vec{R}\right)$ are Slater determinants of single-particle orbitals for spin-up and spin-down electrons. The Jastrow exponent $J\left(\vec{R}\right)$ consisted of electron--electron, electron--ion, and electron--electron--ion polynomials, as well as electron--electron plane-wave expansions, with the coefficients being optimizable parameters \cite{Drummond:04}. The orbitals in the Slater determinants were generated using \textsc{castep} with the Perdew-Burke-Ernzerhof (PBE) \cite{PBE} XC functional and re-represented in a B-spline (blip) basis for use in \textsc{casino} \cite{Hernandez_Gillan_blip:1997,Alfe_Gillan_blip:2004}.

\begin{figure}[htbp!]
    \centering
    \includegraphics[width=0.9\columnwidth]{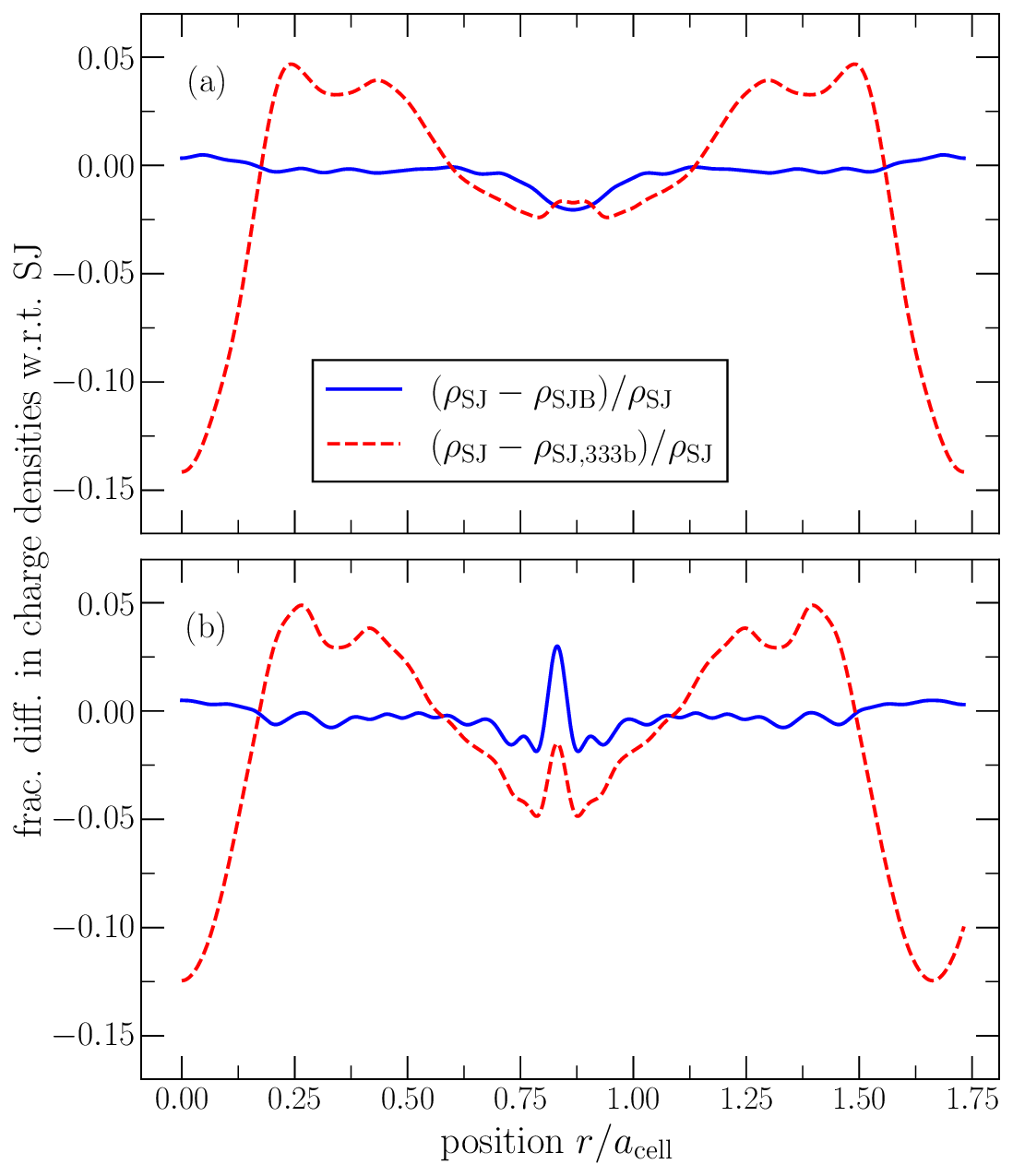}
    \caption{
        Fractional difference between extrapolated estimates of the DMC charge density in a 111b cell with SJB and SJ trial wave functions ($\rho_\text{SJB}$ and $\rho_\text{SJ}$), and between extrapolated estimates of the DMC charge density in 333b and 111b cells with an SJ trial wave function ($\rho_\text{SJ,333b}$ and $\rho_\text{SJ}$). Results are shown for (a) Si and (b) Ge.
        $r$ is the distance along a straight line from the origin through the corner of the conventional unit cell, passing through its center.  $a_\mathrm{cell}$ is the lattice parameter.
    }
    \label{fig:Si_Ge_charge_densities}
\end{figure}

For greater accuracy in fixed-node DMC calculations for a particular system, Slater-Jastrow-backflow (SJB) \cite{Kwon:98,LopezRios:06} wave functions can be used. However, SJB wave functions are much more expensive to evaluate and, as shown in Fig.\ \ref{fig:Si_Ge_charge_densities}, the difference between SJ and SJB charge densities is small compared to finite-size errors at the system sizes for which backflow calculations are feasible. Hence in practice, greater accuracy can be achieved by using an SJ wave function and studying larger simulation supercells. Throughout, the shorthand ``$nnn$b'' has been used to denote a calculation performed in a $n\times n\times n$ simulation supercell with the simulation-cell Bloch k-vector lying at the Baldereschi point (see Sec.\  I of the supplementary material).

Trial wave functions were optimized using two methods. The first was minimization of the variance of the local energy \cite{Umrigar:88,Drummond:05}, which was used to optimize linear parameters in the Jastrow exponent only, providing a good initial wave function. The second was minimization of the energy expectation value \cite{Toulouse:07,Umrigar:07}, which was used to optimize all free parameters and hence provide the final trial wave function.
\subsubsection{Expectation value of the charge density}

The DMC algorithm generates electron configurations distributed as the mixed distribution $|\Psi_\text{T}^*(\vec{R})\psi_\text{DMC}(\vec{R})|$,
where $\psi_\text{DMC}(\vec{R})$ is the fixed-node ground-state wave function.
The charge density operator
\begin{equation}
    \hat{\rho} (\vec{r}) =\sum_{i=1}^{N}\delta\left(\vec{r}-\hat{\vec{r}}_i\right)
\end{equation}
does not commute with the Hamiltonian, so the DMC mixed estimate of the charge density $\rho_\text{DMC}(\vec{r})=\langle \Psi_\text{T}|\hat{\rho}(\vec{r})|\psi_\text{DMC}\rangle/\langle \Psi_\text{T}|\psi_\text{DMC}\rangle$ is not equal to the pure estimate $\langle \psi_\text{DMC}|\hat{\rho}(\vec{r})|\psi_\text{DMC}\rangle/\langle \psi_\text{DMC}|\psi_\text{DMC}\rangle$, unlike the case for the energy expectation value. We obtain a better approximation to the pure charge density using the extrapolated estimation method \cite{Ceperley:1979}. In this approach, we combine the VMC and DMC expectation values to eliminate systematic errors that are first order in the error $\Psi_\text{T}-\psi_\text{DMC}$ in the trial wave function by evaluating the extrapolated DMC charge density as
\begin{equation}
    \rho\left(\vec{r}\right) \approx 2\rho_{\text{DMC}}\left(\vec{r}\right) - \rho_{\text{VMC}}\left(\vec{r}\right),
\end{equation}
where $\rho_{\text{VMC}}(\vec{r})=\langle \Psi_\text{T}|\hat{\rho}(\vec{r})|\Psi_\text{T}\rangle/\langle \Psi_\text{T}|\Psi_\text{T}\rangle$ is the VMC charge density.
The remaining error in the extrapolated density $\rho(\vec{r})$ is second order in the error in the trial wave function.
In each case, two separate DMC calculations were performed, with time steps $\Delta t_1 = 0.04$ Ha$^{-1}$ and $\Delta t_2 = 0.01$ Ha$^{-1}$, and the DMC mixed estimate of the charge density in the limit of zero time step was calculated as
\begin{equation}
    \rho_{\text{DMC}}\left(\vec{r}\right) = \frac{\rho_2\left(\vec{r}\right) - \rho_1\left(\vec{r}\right)}{\Delta t_1 - \Delta t_2}\Delta t_1 + \rho_1\left(\vec{r}\right),
\end{equation}
where $\rho_1$ and $\rho_2$ are the DMC mixed estimates of the charge density at time steps $\Delta t_1$ and $\Delta t_2$, respectively. The DMC target walker population was varied in inverse proportion to the time step.
Further details about the accumulation and pre-processing of densities for inversion is given in Sec.\ II of the supplementary material.

\subsection{Inversion of charge densities}\label{section:inversion_algorithm}
We follow the density inversion method of
Refs.~\onlinecite{Hollins:2017_LFX,Hollins:2017_LFX_metals,den_inv_Callow:2020,den_inv_Sofia_Extension_to_Tim_Callow:2022,Ravindran:2024}
implemented within the \textsc{castep} code.
At the start of the algorithm, the trial KS potential
for spin channel $\sigma$
$\vks^{\sigma (n=0)}$
is initialized to the PBE KS potential using the QMC charge density
$\qmcspinden$,
\begin{equation}
   \begin{split}
        \vks^{\sigma (n=0)}(\vec{r}) = \vext(\vec{r}) + \vhar[\qmcden](\vec{r})
        \\ +
        \vxcs^{\sigma,\mathrm{PBE}}[\qmcden](\vec{r}) ,
   \end{split}
\end{equation}
where each term on the right hand side corresponds to the
external (pseudopotential), Hartree, and PBE exchange-correlation (XC) potentials respectively.

In a steepest descent algorithm the potential $\vks^\sigma(\vec{r})$ at iteration $n$ is updated according to
\begin{equation}
    \vks^{\sigma(n+1)}(\vec{r}) =
    \vks^{\sigma(n)}(\vec{r}) - \epsilon\int\intd{r'} \,
    \frac{\qmcspinden(\vec{r'})-\rho_v^{\sigma(n)}(\vec{r'})}{|\vec{r}-\vec{r'}|}.
\end{equation}
Here, $\rho_v^{\sigma(n)}$ is the density of $\vks^{\sigma(n)}(\vec{r})$ at iteration $n$ calculated via
\begin{equation}
    \rho_v^{\sigma(n)}(\vec{r})  = \sum_i^\mathrm{occ} \sum_\vec{k} f_{i\vec{k}} w_\vec{k} |\phi_{i\vec{k}}^\sigma(\vec{r})|^2 .
\end{equation}
The parameter $\epsilon >0 $ controls the rate of descent. %
The orbitals at each band $i$ and k-point $\vec{k}$, $\{\phi^\sigma_{i\vec{k}}\}$, and occupancies $f_{i\vec{k}}$ and weights $w_\vec{k}$
are obtained by solving the (spin-)KS equations
\begin{equation}
    \left(-\frac{1}{2}\nabla^2 + \vks^{\sigma(n)}(\vec{r})\right)\phi_{i\vec{k}}^{\sigma(n)}(\vec{r})
    =
    \varepsilon_{i\vec{k}}\phi_{i\vec{k}}^{\sigma(n)}(\vec{r}).
\end{equation}
This procedure is repeated until the Coulomb energy
\begin{equation}
    U
    =
    \sum_\sigma
    \frac{1}{2}\iint\intd{r} \, \intd{r'} \,
    \frac{
        [\Delta\rho^{\sigma(n)}(\vec{r})]
        [\Delta\rho^{\sigma(n)}(\vec{r'})]
    }
    {|\vec{r}-\vec{r'}|}\geq 0,
    \label{eq:LFX_obj_func}
\end{equation}
associated with the difference between the two densities,
$\Delta \rho^{\sigma (n)}(\vec{r}) = \qmcspinden(\vec{r})-\rho_v^{\sigma (n)}(\vec{r})$,
is sufficiently minimized.

In practice, the correction to the potential is done using the more efficient Fletcher-Reeves conjugate gradient algorithm \cite{Fletcher_Reeves_CG}, where the optimal value of $\epsilon$ is found using a parabolic line search to accelerate convergence.
The Coulomb energy $U$ is monitored over a set of four iterations and is deemed to be converged when the difference between the maximum and minimum value within this set was less than $10^{-8}$ Ha/atom.
Note that, for consistency in the inversion of QMC densities in \textsc{castep}, we used identical plane-wave cutoffs and $k$-points to the original DFT calculations used to generate the Slater determinants for the QMC calculations.

\section{Errors in the charge density \label{section:errors}}

\begin{table*}[htbp!]
 \caption{
  Calculated KS band gaps via inversion of the QMC density with various supercell sizes.
  The unsymmetrized (unsym) columns contain gaps obtained from inversion \textit{without} explicit imposition of symmetry on the charge density.
  The symmetrized (sym) results are obtained via the appropriate imposition of symmetries on the charge density according to the symmetry operations of the crystallographic space group of the given material.
  Gaps are extrapolated (from unsymmetrized results) to infinite cell size according to Eq.\ \eqref{eq:gap_fit}.
  Experimental lattice parameters from Ref.\ \onlinecite{Madelung:1996} were used unless otherwise stated.
  }
 \label{table:supercell_gaps}
 \begin{ruledtabular}
  \begin{tabular}{ld{3}d{3}d{3}d{3}d{3}d{3}d{3}d{3}d{3}}
   {}& \multicolumn{8}{c}{KS band gap (eV)} & {}\\% & {} \\
   Material & \multicolumn{2}{c}{111b cell}&
   \multicolumn{2}{c}{222b cell}&
   \multicolumn{2}{c}{333b cell}&
   \multicolumn{2}{c}{444b cell}
   & \multicolumn{1}{r}{Extrapolated}%
   \\
   {}& \multicolumn{1}{c}{unsym}& \multicolumn{1}{c}{sym}&
   \multicolumn{1}{c}{unsym}& \multicolumn{1}{c}{sym}&
   \multicolumn{1}{c}{unsym}& \multicolumn{1}{c}{sym}&
   \multicolumn{1}{c}{unsym}& \multicolumn{1}{c}{sym}
   &{} %
   \\
   \hline
   Si& 0.550& 0.627& 0.713& 0.748& 0.783& 0.798& 0.799& 0.815&
   0.812\\
   Diamond & 3.993& 4.125& 4.264& 4.345& 4.370& 4.400&
   \multicolumn{2}{c}{-}& 4.414\\
   GaAs& 1.115& 1.212& 1.293& 1.309& 1.346& 1.354&
   \multicolumn{2}{c}{-}& 1.368\\
   Ge& 0.123& 0.267& 0.315& 0.334& 0.360& 0.366&
   \multicolumn{2}{c}{-}& 0.379\\
   NaCl& 4.957& 4.952& 5.562& 5.576& 5.641& 5.648&
   \multicolumn{2}{c}{-}& 5.674\\
   BaTiO\textsubscript{3}\footnote{Lattice parameter from Refs.\ \onlinecite{BaTiO3_Lat_Param_1} and \onlinecite{BaTiO3_Lat_Param_2}}&  3.390& 3.553& 3.252& 3.288& 3.219& 3.242&
   \multicolumn{2}{c}{-}& 3.206\\
   SrTiO\textsubscript{3}\footnote{Lattice parameters from Ref.\ \onlinecite{SrTiO3_Lat_Param}}& 3.604& 3.719& 3.454& 3.471& 3.432& 3.469& \multicolumn{2}{c}{-}& 3.422\\
   MnO\footnote{Lattice parameters from Ref.\ \onlinecite{TMO_lat_params}}& 2.570& 2.595& 2.914& 2.914& 2.943& 2.952&
   \multicolumn{2}{c}{-}& 2.952\\
  \end{tabular}
 \end{ruledtabular}
\end{table*}

\subsection{Finite size errors}%
\label{section:results:extrapolation}
Finite-size effects are a source of systematic error in explicitly correlated methods such as QMC\@.

\begin{figure}[htbp!]
    \centering
    \includegraphics[width=\columnwidth]{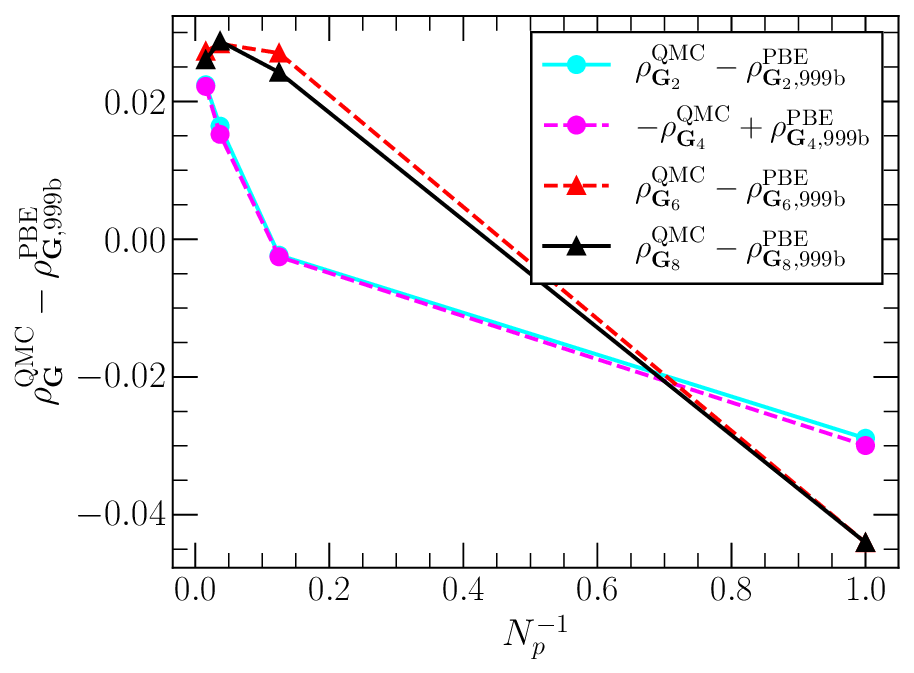}
    \caption{
        Difference between the $n$th Fourier component of the density of Si as evaluated by QMC in a supercell containing $N_\text{p}$ primitive cells with the supercell Bloch vector at the Baldereschi MVP ($\qmcdeng{n}$) and the $n$th Fourier component of the density as evaluated by PBE using a $9\times 9\times 9$ k-point grid centered on the Baldereschi MVP ($\rho_{\vec{G},999\text{b}}^\text{PBE}$), plotted against the reciprocal of $N_\text{p}$.
        Note that $\qmcdeng{2}$ and $\qmcdeng{4}$ are symmetry equivalent, as are $\qmcdeng{6}$ and $\qmcdeng{8}$.
        The differences between the symmetry equivalent Fourier coefficients are indicative of the random errors in the QMC results.
        }
    \label{fig:rho_fourier_finite_size}
\end{figure}

In Fig.\ \ref{fig:rho_fourier_finite_size}, we plot the differences between QMC- and PBE-calculated Fourier coefficients of the density in Si. The QMC results are obtained in different supercells, while the PBE results are obtained using a fine k-point grid.
Random errors in the QMC results are small compared to finite-size errors, as shown by the similarity of symmetry-equivalent Fourier coefficients at each system size.
Furthermore, for sufficiently large supercells it is clear that finite-size errors in the Fourier coefficients are small compared with the difference between the QMC and PBE results.

\begin{figure}[htbp!]
    \centering
    \includegraphics[width=\columnwidth]{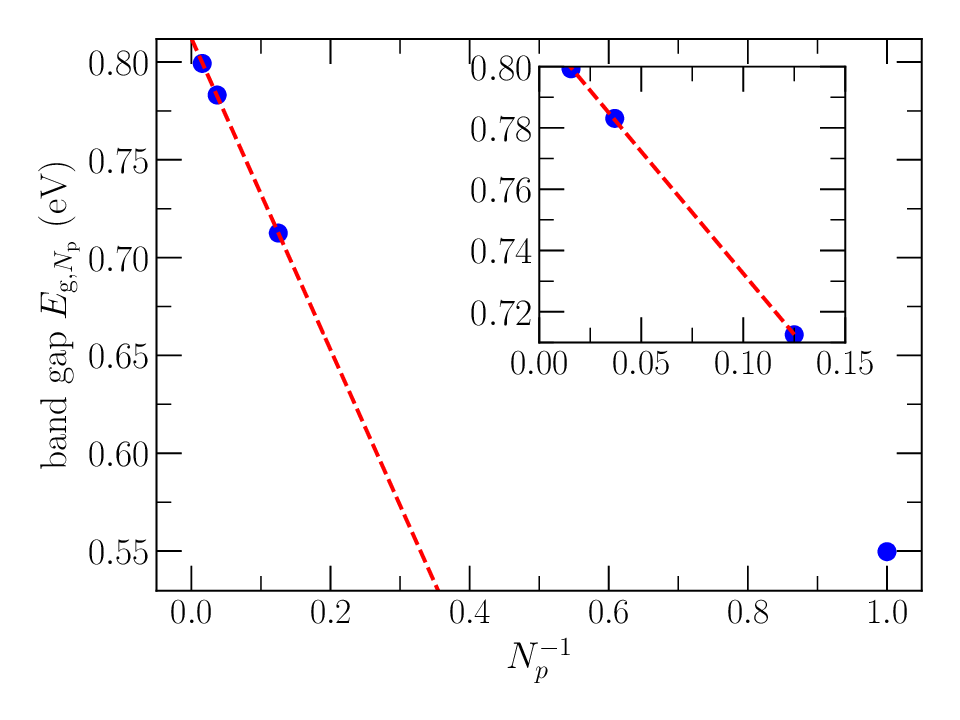}
    \caption{
        KS band gaps $E_{\mathrm{g},N}$ calculated via inversion of the Si density  obtained from various QMC simulation supercells consisting of $N_\text{p}$ primitive cells (equivalently, the number of k-points within the inversion/DFT calculation).
        The dashed red line shows a linear fit of the gap $E_{\mathrm{g},N}$ as a function of $N_\text{p}^{-1}$, yielding an extrapolated band gap of 0.81\ eV at infinite system size.
    }
    \label{fig:Si_extrap_gap}
\end{figure}

Systematic finite-size effects in the total energy per primitive cell scale asymptotically as $\mathcal{O}(N_\text{p}^{-1})$ \cite{Chiesa:06}. Therefore, the total energy per primitive cell can be extrapolated to infinite system size according to
\begin{equation}
	E_{N_\text{p}} = E_\infty + \frac{c}{N_\text{p}},
	\label{eq:lin_fit}
\end{equation}
where $\{E_{N_\text{p}}\}$ is the total energy per primitive cell obtained from a computational cell consisting of $N_\text{p}$ primitive cells, $c$ is a fitting parameter and $E_\infty$ is the energy per primitive cell in the thermodynamic limit. We excluded ``supercells'' consisting of a single primitive cell from the fit.
We performed a similar fit for the KS gaps $\{E_{\mathrm{g},N_\text{p}}\}$ obtained from KS potentials $\vks(\vec{r})$ by inversion:
\begin{equation}
	E_{\mathrm{g},N_\text{p}} = E_{\mathrm{g},\infty} + \frac{\tilde{c}}{N_\text{p}},
	\label{eq:gap_fit}
\end{equation}
where $\tilde{c}$ is a fitting parameter and $E_{\mathrm{g},\infty}$ is the extrapolated KS gap.
Note that, as mentioned in Sec.\ \ref{section:inversion_algorithm}, if the QMC density is generated in a supercell consisting of $N_\text{p}$ primitive cells, the inversion calculation is performed using the commensurate Monkhorst-Pack grid of $N_\text{p}$ k-points, which was also used for the generation of the Slater determinant in the QMC trial wave function.
Furthermore, since the Monkhorst-Pack grid is not centered on the $\Gamma$-point but on
the Baldereschi MVP, we do not impose time-reversal symmetry on the Monkhorst-Pack grid.

\begin{figure}[htbp!]
    \centering
    \includegraphics[width=\columnwidth]{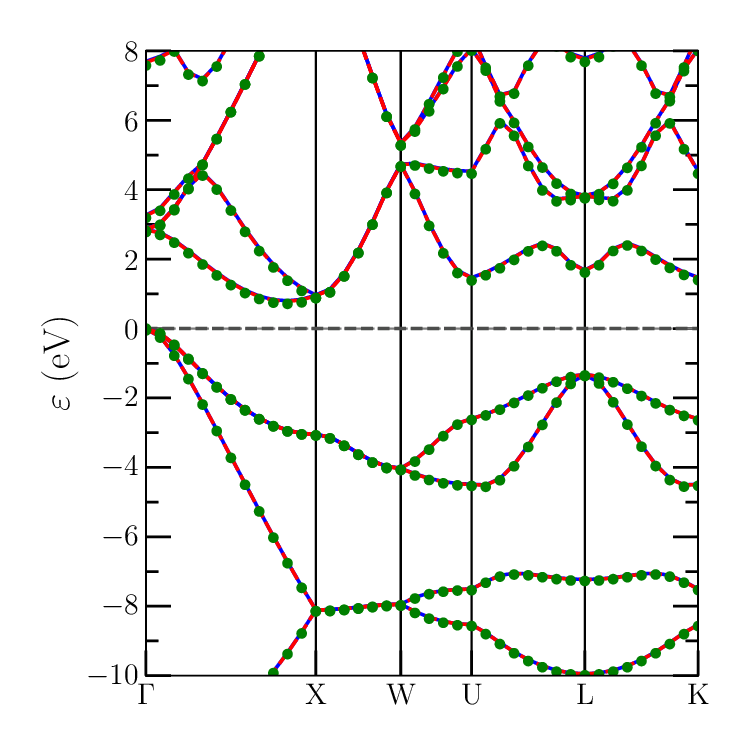}
    \caption{Calculated KS band structure for Si using the inverted density from various supercells. The color scheme is as follows: blue 444b supercell, red 333b supercell, green 222b supercell.}
    \label{fig:Si_bs_fin_size}
\end{figure}

The gaps for each supercell and system studied are given in Table \ref{table:supercell_gaps} along with the extrapolated KS gap.
We show an example of this extrapolation for Si in Fig.\ \ref{fig:Si_extrap_gap}, where the extrapolated KS band gap is 0.81 eV\@, while in Fig.\ \ref{fig:Si_bs_fin_size} we show the calculated KS band structure obtained via inversion of the QMC density from a computational supercell consisting of $N_\text{p}$ primitive cells.
As a general rule, calculations performed in the primitive cell are dominated by finite-size errors and thus are outliers. However, one can see that the KS gaps $E_{\mathrm{g},N}$ obtained from the 222b, 333b, and 444b supercells vary linearly with $N_\text{p}^{-1}$.
We assume this linear scheme holds for the other systems studied such that $E_{\mathrm{g},\infty}$ can be obtained via extrapolation from 222b and 333b cells.

\subsection{Statistical errors}

Ideally, the random errors of the simulation should be considerably smaller than the systematic finite-size errors.
The random errors in the Fourier components of the density are correlated. In principle, one could calculate the covariance matrix between the Fourier components $\rho_\vec{G}$. However, given that the number of Fourier components is of the order of tens of thousands, the calculation of the covariance matrix would be computationally demanding, and the propagation of statistical errors to the density-dependent KS gap would also be very challenging, as it would require partial derivatives of the gap with respect to each $\rho_\vec{G}$.
In practice, the simplest, albeit computationally demanding, approach for obtaining error bars on quantities derived from the charge density, such as the inverted-density KS band gap, is to repeat the entire calculation several times and take the mean and the standard error in the mean of the results obtained.

Alternatively, one can obtain a rough estimate of the random errors in the charge density by comparing Fourier components $\rho_\vec{G}$ that ought to be identical under the symmetry of the crystal, as done in Fig.\ \ref{fig:rho_fourier_finite_size}.
Due to the stochastic nature of the Monte Carlo techniques, no symmetries are explicitly imposed on the charge density during VMC and DMC calculations.
However, any expected symmetries in the charge density should still arise up to statistical error. By examining the charge density Fourier components. $\rho_\vec{G}$, one can instead verify that the expected symmetries in the densities are present up to statistical error.

Therefore, as a preliminary step, one can ensure that the Fourier components of the charge density $\rho_\vec{G}$ satisfy the expected symmetries of a given system up to a certain numerical threshold. In particular, since the density $\rho(\vec{r})$ is a real function, its Fourier components satisfy $\rho_\vec{G}^* = \rho_\vec{-G}$.
Furthermore, when the crystal possesses inversion symmetry, this property is shared by the density.
In reciprocal space, the Fourier components thus satisfy
$\rho_\vec{-G} = \rho_\vec{G}$.
This identity, along with the previous one for the Fourier components of a real function implies that the Fourier components themselves must be real, satisfying the relation
 \begin{equation}
    \rho_\vec{G}^* = \rho_\vec{-G} = \rho_\vec{G}.
\end{equation}
To check whether inversion symmetry is satisfied, we compare Fourier components at reciprocal lattice vectors $\pm \vec{G}$ according to
\begin{equation}
    |\rho_\vec{G}-\rho_\vec{-G}| \leq \epsilon_\mathrm{atol} + \epsilon_\mathrm{rtol} \times
    | \rho_\vec{-G} |,
\end{equation}
where $\epsilon_\mathrm{atol}$ is the absolute tolerance and $\epsilon_\mathrm{rtol}$ is the relative tolerance.
For systems with inversion symmetry, this inequality was satisfied with $\epsilon_\mathrm{atol}=10^{-6}$ a.u.\ and $\epsilon_\mathrm{rtol}=10^{-5}$.

Furthermore, since the input target density for the inversion is ultimately done on a real space rectilinear grid, the real space density $\qmcden(\vec{r})$ can be checked by symmetrizing it according to the symmetry operations of the corresponding space group.

The integrated absolute difference (IAD) between the symmetrized $\qmcden^\mathrm{sym}(\vec{r})$ and unsymmetrized density $\qmcden(\vec{r})$ is defined as
\begin{equation}
    \mathrm{IAD} = \frac{1}{N_\text{r}}\sum_{n=1}^{N_\text{r}}|\rho_\mathrm{QMC}(\vec{r}_n)-\rho_\mathrm{QMC}^\mathrm{sym}(\vec{r}_n)|,
    \label{eq:IAD}
\end{equation}
where $N_\text{r}$ is the number of real-space grid points $\vec{r}_n$.
We found that the mean IAD per electron averaged over all materials and simulation cells was $\sim 0.022$.

To further check the KS potentials, we performed inversion of the symmetrized density $\rho_\mathrm{QMC}^\mathrm{sym}$ and compared the calculated KS band structures with the band structures obtained from inversion of the density without the explicit imposition of symmetry. Note that the running density $\rho_v(\vec{r})$ (i.e., density for the potential at each iteration)
in this inversion calculation is also symmetrized along with the KS potential $\vks(\vec{r})$ at each iteration. The Coulomb energy [see Eq.\ \eqref{eq:LFX_obj_func}] is likewise calculated from both the symmetrized $\qmcden(\vec{r})$ and $\rho_v(\vec{r})$.
The calculated KS gaps from the inversion of the symmetrized QMC density $\rho_\mathrm{QMC}^\mathrm{sym}(\vec{r})$ and unsymmetrized density $\qmcden(\vec{r})$ are given in Table\ \ref{table:supercell_gaps}.
The difference between the band structures can be quantified using the mean absolute eigenvalue difference (MAED)
\begin{equation}
    \mathrm{MAED} = \frac{1}{N_b  N_k}\sum_{i,\vec{k}}
    |\varepsilon_{i\vec{k}}-\varepsilon_{i\vec{k}}^\mathrm{sym}|,
\end{equation}
where $N_b$ and $N_k$ are the numbers of bands and k-points, respectively, in the band structure and the $\{\varepsilon_{i\vec{k}}^\mathrm{sym}\}$ are the eigenvalues of the KS Hamiltonian with potential $\vks(\vec{r})$ obtained from $\qmcden^\mathrm{sym}(\vec{r})$.
Naturally, if running a spin polarized calculation, we average over all spins as well with the denominator in the prefactor changing from $N_b N_k \to 2 N_b N_k$.
We found the MAED to be of the order of $10^{-2}$ eV with the averaged MAED over all systems and computational supercells being $4\times10^{-2}$ eV\@.

We observed a larger deviation between the symmetrized and unsymmetrized QMC densities obtained from a single primitive cell.
As previously discussed, the error in this density also has a contribution from systematic finite-size effects which are particularly significant for calculations in a single primitive cell (see Table\ \ref{table:supercell_gaps}).
If we exclude these densities from the calculation of the mean IAD, we find that the mean IAD per electron is reduced to $\sim 0.01$ while the MAED is reduced to $2\times10^{-2}$ eV\@. The small difference between the symmetrized and unsymmetrized calculations, particularly for QMC densities in larger supercells, is encouraging, suggesting that
the random error in the calculation is small and that finite-size error is more significant.

\begin{figure*}[!htbp]
    \includegraphics[width=0.85\linewidth]{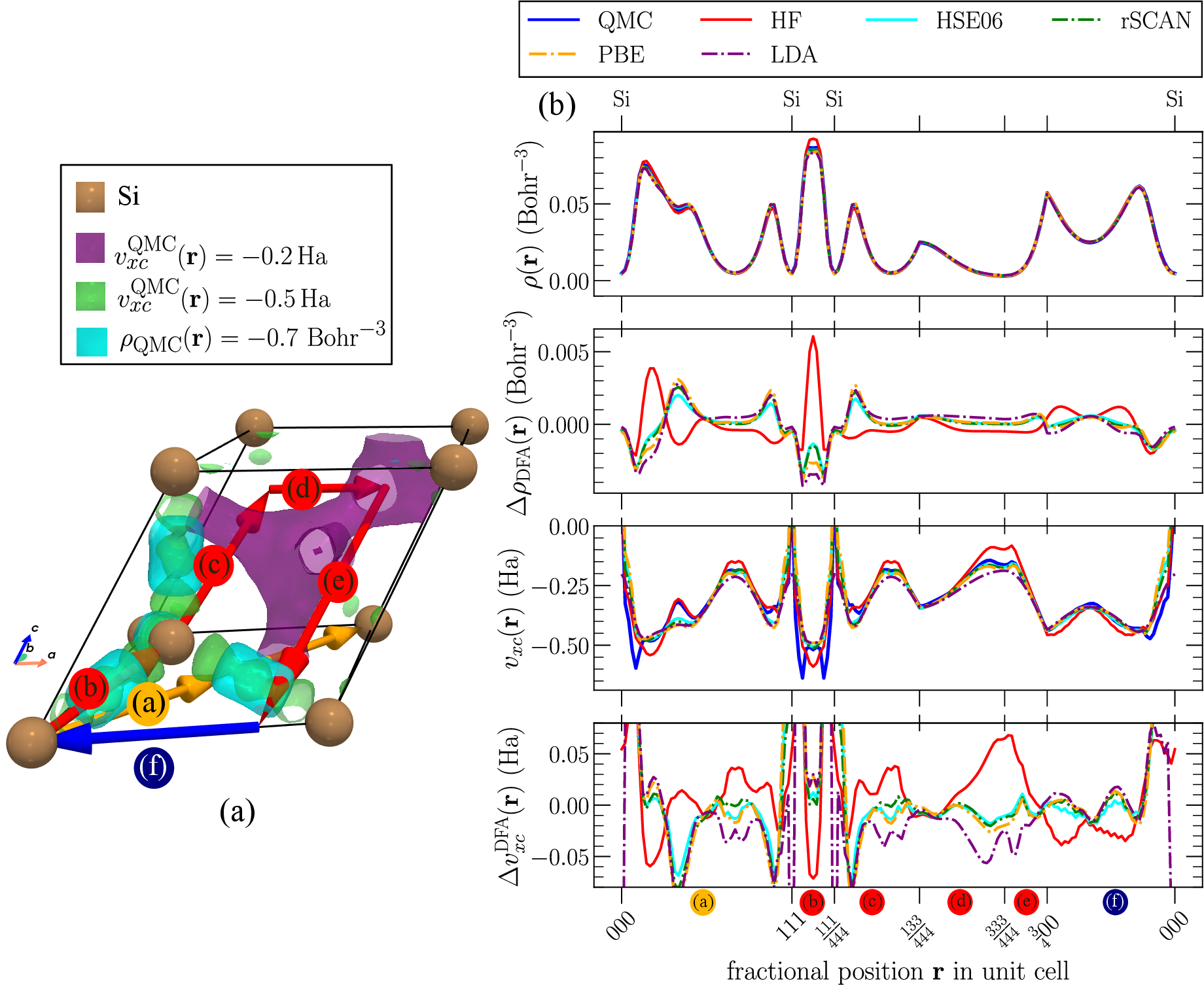}
    \caption{
        (a) Isosurfaces of the (unsymmetrized) QMC density $\qmcden(\vec{r})$ and $\vxcs^\mathrm{QMC}(\vec{r})$ in bulk Si.
        (b) In the first and third panels, the density of each method $\rho(\vec{r})$ and the corresponding $\vxcs(\vec{r})$, respectively, are plotted along the path through the unit cell shown by the coloured arrows in (a) (the colours of the arrows are help distinguish portions of the path). This respective portion of the path is indicated by the coloured circles in (a) and in the x-axis of (b).
        In the second and fourth panels, we plot the density difference $\Delta\rho_\mathrm{DFA}(\vec{r})$ and the XC potential difference $\Delta \vxcs^\mathrm{DFA}(\vec{r})$ from the respective QMC results for various DFAs.
        The isosurface for the density has been chosen to coincide with regions of high electron density associated with bonding.
    }
    \label{fig:Si_locpot_compare}
\end{figure*}

\section{Assessing densities and local potentials} \label{section:exact_KS}
\subsection{Behavior of exchange-correlation potentials}
Using the QMC density obtained with the largest computational supercell as a benchmark, we now assess
the quality of densities and XC potentials from various density functional approximations (DFAs) at
different rungs of Jacob's ladder \cite{Jacobs_ladder}, namely the LDA, PBE [a generalized gradient approximation (GGA)], regularized SCAN (rSCAN; a meta-GGA), and HSE06 (a hybrid DFA), in addition to Hartree-Fock (HF)\@.
We note that meta-GGAs, hybrid DFAs, and HF are explicit functionals of the single-particle orbitals and are thus nonlocal, implicit functionals of the density; in hybrids and HF, the nonlocality arises from the Fock exchange term and in meta-GGAs due to terms involving the kinetic energy density.
Such DFAs can be treated either in a KS scheme via the OEP method \cite{Sharp_Horton:1953_OEP,Talman_Shadwick_OEP_1976,Kuemmel_OEP_Review_2008},
or using a generalized Kohn-Sham (GKS) scheme \cite{band_gaps_in_GKS_2017,GKS_OEP_SCAN_Perdew_2016} with a nonlocal effective potential.

For these nonlocal DFAs, we note that a KS potential $\vks(\vec{r})$ can also be obtained by inverting the GKS target density $\rho_\mathrm{GKS}(\vec{r})$.
In particular, as pointed out in Refs.\ \onlinecite{Hollins:2017_LFX}, \onlinecite{Hollins:2017_LFX_metals}, and \onlinecite{Ravindran:2024}, the exchange-only potential obtained by inversion of the HF density, dubbed the local Fock exchange (LFX) potential $v_x^\mathrm{LFX}(\vec{r})$ in these references, is similar to the exchange-only OEP\@.
Furthermore, it is expected that the XC potential $\vxcs(\vec{r})$ [referred to as the local exchange-correlation (LXC) potential in Ref.\ \onlinecite{Ravindran:2024}] obtained from the inversion of any GKS density will be similar to that obtained via the OEP method (except possibly in strongly correlated systems).
A thorough discussion and comparison of GKS, OEP, and KS inversion calculations
(referred to as inv-OEP) can be found in Ref.~\onlinecite{Weitao_Yang_Faraday:2020}.

In order to facilitate a like-for-like comparison, it is crucial that the external potential for each calculation is identical, so that any differences arise solely due to the DFA's XC functional. Therefore, to ensure consistency, all calculations were carried out using the same TN pseudopotential (including the same local angular momentum channel).

\begin{figure*}[!htbp]
    \includegraphics[width=0.85\linewidth]{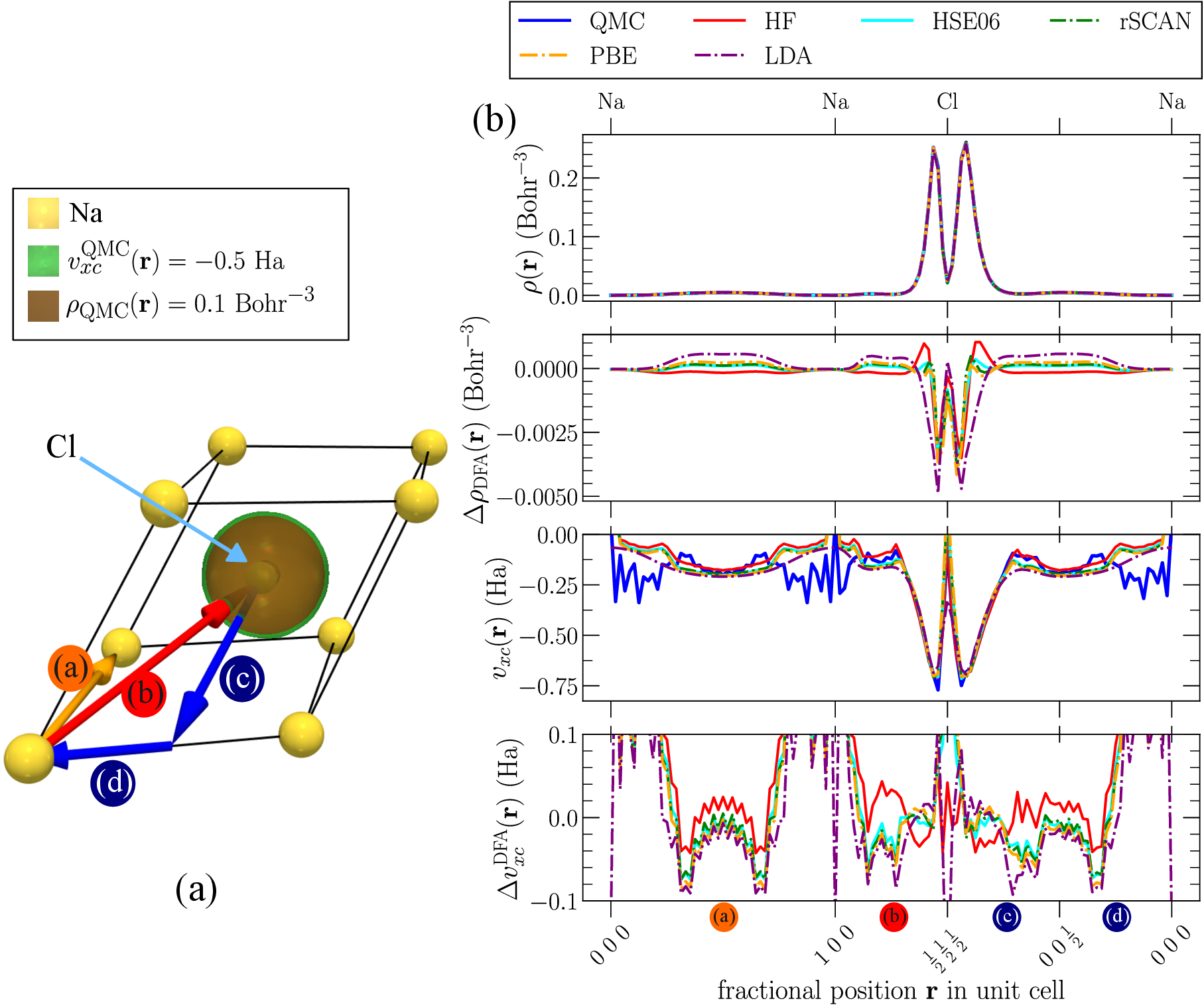}
    \caption{
        Same as Fig.\ \ref{fig:Si_locpot_compare} for NaCl.
        Note that the Cl atom at the center of the cell is marked by the blue arrow. Isosurfaces have been chosen to coincide with regions of high electron density associated with bonding.
    }
    \label{fig:NaCl_locpot_compare}
\end{figure*}

In Fig.\ \ref{fig:Si_locpot_compare}, we compare the densities of each method along with the (local) $\vxcs(\vec{r})$ in Si. The density is plotted along a path through the unit cell shown by the red arrows in Fig.\ \ref{fig:Si_locpot_compare}(a).
As one might expect, the potential is deeper (more negative) where the density is higher, for instance in the bonding region as highlighted by the cyan and green isosurfaces in Fig.\ \ref{fig:Si_locpot_compare}(a).
Moreover, it can be seen that the HF density is overlocalized compared to the QMC density, particularly in the bonding region along the bond axis from the Si atom at $(1,1,1) = (0,0,0)$ to the other Si atom at $(1/4,1/4,1/4)$.
The LFX potential obtained from the inversion of the HF density is consequently too deep. By contrast, the other DFAs have a slight delocalization error in their densities and correspondingly shallower $\vxcs(\vec{r})$ than $\vxcs^\mathrm{QMC}(\vec{r})$ in this region.
The spikes observed in the XC potential near ions appear to be
a pseudopotential artifact due to the valence charge density tending to zero
inside the core region of the pseudopotential.
As discussed in Appendix \ref{appendix:nlcc_effects} for the PBE potential
$\vxcs^\mathrm{PBE}(\vec{r})$, this effect arises even in standard DFT calculation using the latest \textsc{castep} on-the-fly norm-conserving (NCP19) pseudopotentials \cite{Lin_CASTEP_NCP:1993,CASTEP_NCP19_Reference} when nonlinear core corrections are
\textit{not included}; nonlinear core corrections are also absent in the TN pseudopotentials.
Similar spikes have been observed in the inversion calculations of auxiliary field QMC densities by Aouina and coworkers in Refs.\ \onlinecite{den_inv_Aouina_Reining_QMC:2023} and \onlinecite{den_inv_Aouina_QMC:2024}, which used
optimized norm-conserving pseudopotentials (ONCVPSP) without nonlinear core corrections generated according to the method of Hamann \cite{Hamann_ONCVPSP:2013}.

\begin{turnpage}
    \begin{figure*}[htbp!]
    \includegraphics[width=\linewidth]{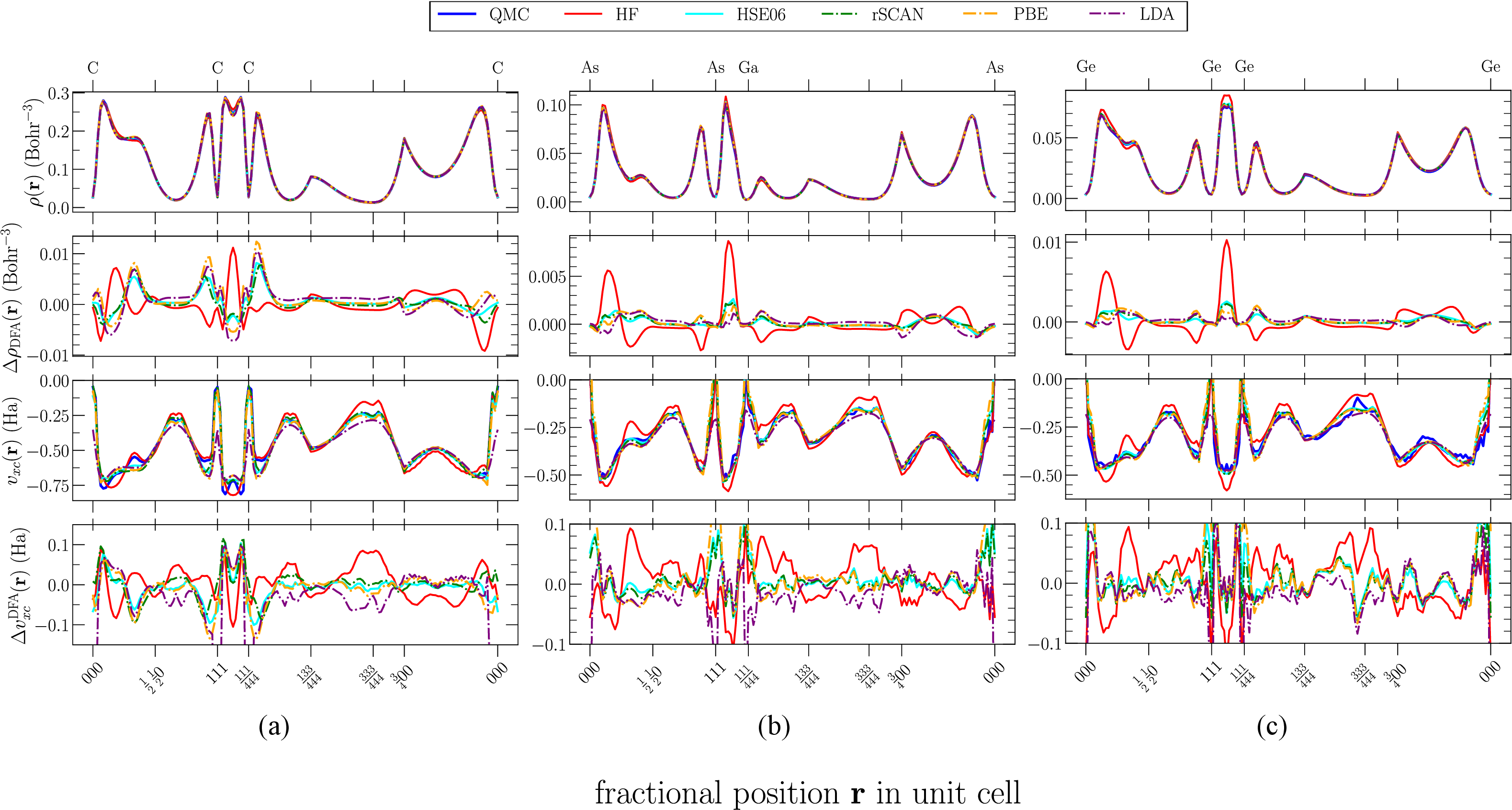}
    \caption{
        Same as Fig.\ \ref{fig:Si_locpot_compare}(b) for bulk (a) diamond, (b) GaAs, and (c) Ge. Note that the same path has been used as for bulk Si, shown in Fig.\ \ref{fig:Si_locpot_compare}(a).
    }
    \label{fig:other_zincblende_locpot_compare}
    \end{figure*}
\end{turnpage}

We observed similar behavior in other diamond-like crystal structure semiconductors, namely diamond, GaAs, and Ge as shown in Fig.\ \ref{fig:other_zincblende_locpot_compare}.
It turns out however that the results for GaAs and Ge have a significant systematic error due to the inclusion of atomic semicore electronic states within the pseudopotential, treating them
as core states.
We discuss this in further detail specifically for the inversion of HF densities in Appendix \ref{appendix:pspot_core_effects}.
Nonetheless, the analysis in our work here remains valid given that we have used the same pseudopotential (external potential).
Plots of $v_{Hxc}(\vec{r})$ are provided in the supplementary material for Si and NaCl in Figs.~S8 and S9.

\begin{figure*}[t]
 	\includegraphics[width=0.85\linewidth]{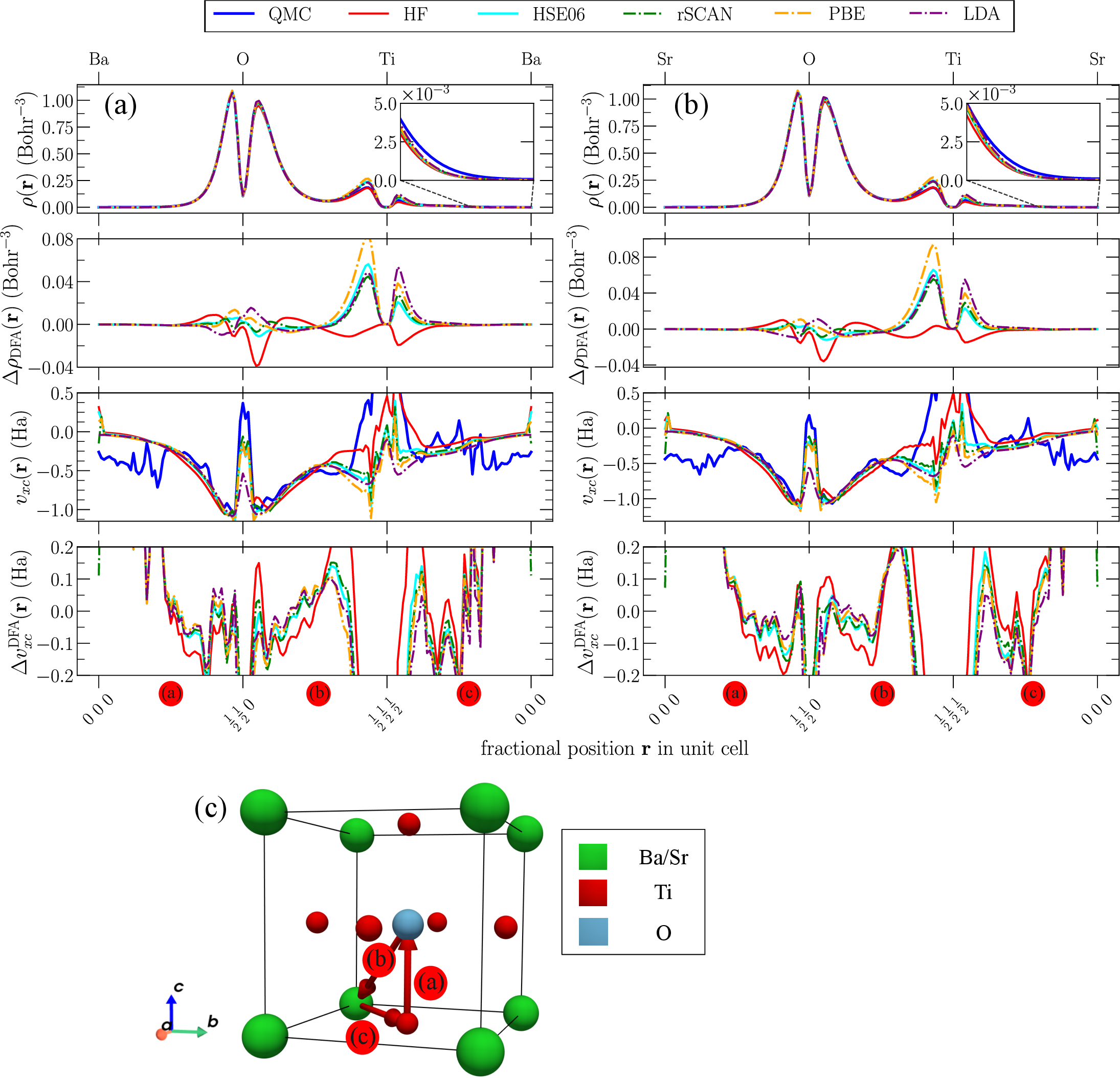}
	\caption{
		Top panel: Comparison of QMC $\rho_\mathrm{QMC}(\vec{r})$ and DFA $\rho_\mathrm{DFA}(\vec{r})$ densities along a path through the unit cell for (a) BaTiO\textsubscript{3} and (b) SrTiO\textsubscript{3}; second panel: density difference of each DFA along the path $\Delta\rho_\mathrm{DFA}(\vec{r})$; third panel: corresponding $\vxcs(\vec{r})$ for QMC and each DFA; and fourth panel: difference in XC potential $\Delta\vxcs^\mathrm{DFA}(\vec{r})$ from the $\vxcs^\mathrm{QMC}(\vec{r})$.
		The path taken through the unit cell is shown in (c). The insets show the density in low density region.
	}
	\label{fig:BTO_STO_locpot_compare}
\end{figure*}

\begin{figure*}[t]
	\includegraphics[clip,width=0.95\textwidth]{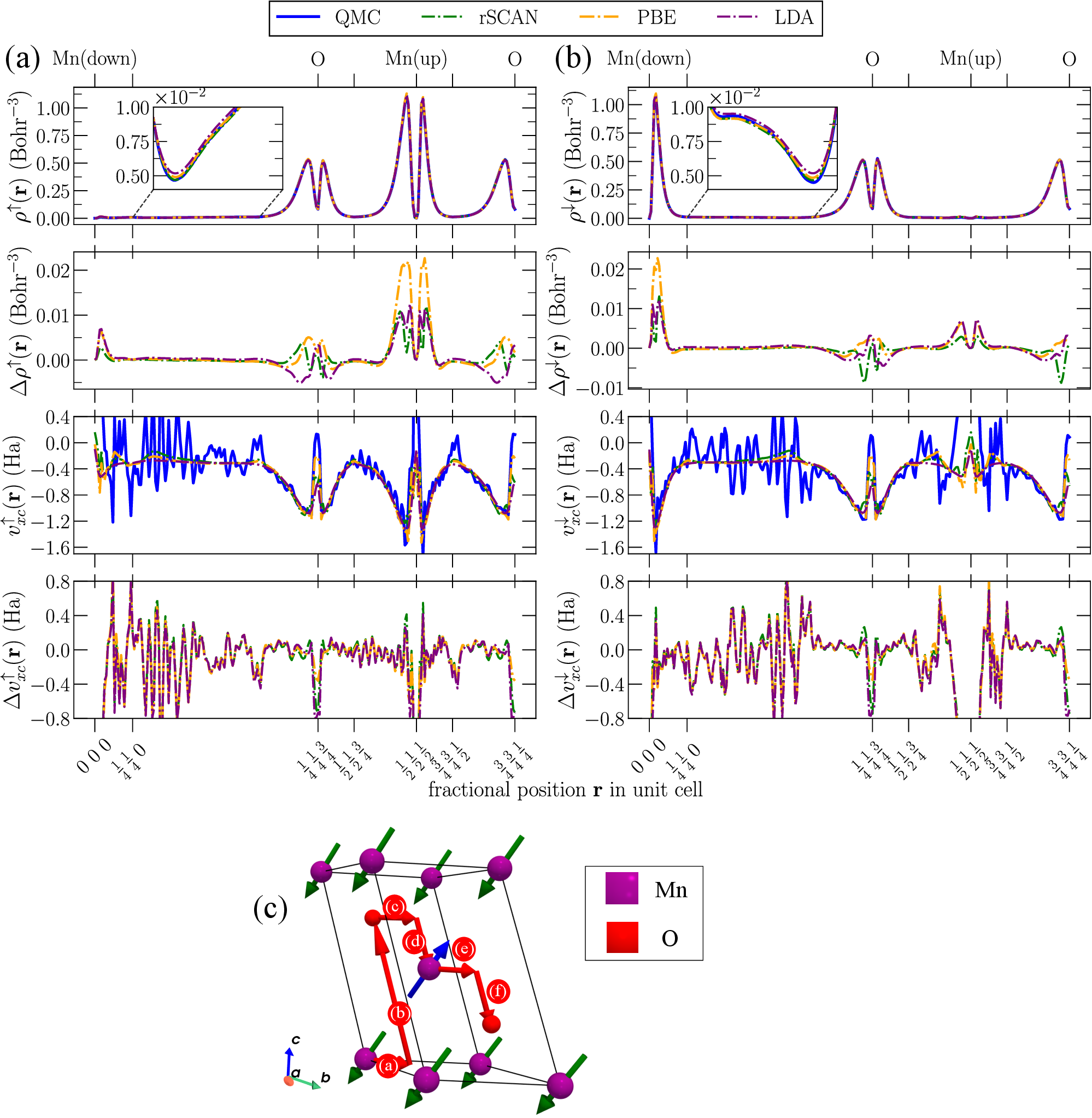}
	\caption{
		Same as Fig.\ \ref{fig:Si_locpot_compare}(b) for bulk MnO\@. The panels on the left show (a) spin-up quantities while the panels on the right are for (b) spin down.
		The path taken through the unit cell is shown in (c).
	}
	\label{fig:MnO_locpot_compare}
\end{figure*}

In systems where the (pseudovalence) charge density is very low (nearly zero) over extended
spatial regions, such as in ionically bonded NaCl, the QMC KS
potential $\vxcs^\mathrm{QMC}(\vec{r})$ undergoes oscillations as shown in
Fig.~\ref{fig:NaCl_locpot_compare}.
On the other hand, $\vxcs^\mathrm{QMC}(\vec{r})$ is well-behaved elsewhere and we note that similar behavior was observed in Ref.~\onlinecite{den_inv_Aouina_QMC:2024}.
For a density that is strictly zero in a region, or even not strictly positive everywhere in the region, the Hohenberg-Kohn (HK) theorem
[i.e.\ the one-to-one correspondence between $\vks(\vec{r})$ and the ground state density $\rho(\vec{r})$] breaks down inside the region, since the
HK theorem requires a density that is nonzero almost everywhere.
A similar result is obtained in the inversion of BaTiO\textsubscript{3} and SrTiO\textsubscript{3} as shown in Figs.~\ref{fig:BTO_STO_locpot_compare}(a) and (b) respectively.
A further discussion of this point can be found in
Refs.~\onlinecite{Capelle_VIgnale_SDFT:2001,Gidopoulos_SDFT:2007,Penz_HK_review_part_I:2023,Penz_HK_review_part_II:2023}.
In our case, the density is not strictly zero in this region but numerically very small and the convergence of the inversion algorithm in such regions is slow and can pose numerical challenges. This is demonstrated and discussed in Sec.\ III of the supplementary material.
Nonetheless, as can be seen in the inset of Fig.~\ref{fig:NaCl_locpot_compare}, the expected qualitative behavior is still observed in these low-density regions, where $\vxcs^{\rm QMC}(\vec{r})$ is deeper than the DFA XC potentials when
the QMC density is slightly larger than the DFAs' densities.
This is also a good demonstration that even small differences in the density can lead to large differences in the KS potential \cite{Savin:2001,Savin_AC_Perspec:2003}.

In MnO, which exhibits antiferromagnetic ordering, we observed similar difficulties in converging the potential where the charge density is low, as shown in Fig.\ \ref{fig:MnO_locpot_compare}. For instance, near the Mn spin-down ion, $\vxcs^\uparrow(\vec{r})$ exhibited severe oscillations [and similarly for $\vxcs^\downarrow(\vec{r})$ near the Mn spin-up ion].
Outside of these regions, however, some more meaningful insight can be drawn from the
behavior of the potential and the density.
In particular, all DFAs considered in this work (with the TN pseudopotentials) overlocalize the electrons around the ions compared to the QMC density, leading to deeper potentials (as expected).
Unfortunately, due to the hardness of the Mn pseudopotential and consequently the necessity for a relatively high plane-wave cutoff energy to converge the basis set, we were unable to run HF or HSE06 calculations for MnO\@.

\begin{table*}[htbp!]
    \caption{Comparison of densities and corresponding Kohn-Sham potentials using the integrated absolute error (IAE) per electron and the energy difference $\mathcal{E}$ per electron (in mHa) as defined in Eq.\ \eqref{eq:HK_metric}.
	The density used as the QMC benchmark is taken from the calculation using the largest computational supercell,
	}
	\begin{ruledtabular}
		\begin{tabular}{ld{2}d{2}d{2}d{2}d{2}cd{3}d{3}d{3}d{3}d{3}}
			\multirow{2}{*}{Material}&\multicolumn{5}{c}{IAE per electron $\times 10^{-2}$}
			& &\multicolumn{5}{c}{$\mathcal{E}$ (mHa)}
			\\
			& \multicolumn{1}{r}{LDA}& \multicolumn{1}{r}{PBE}& \multicolumn{1}{r}{rSCAN}& \multicolumn{1}{r}{HSE06}& \multicolumn{1}{r}{HF}& &
			\multicolumn{1}{r}{LDA}& \multicolumn{1}{r}{PBE}& \multicolumn{1}{r}{rSCAN}& \multicolumn{1}{r}{HSE06}& \multicolumn{1}{r}{HF}
			\\
			\hline
			Si&     2.45&     1.86&     1.50&     1.66&     2.82& &
			0.635&    0.667&    0.488&    0.399&    0.779\\
			Diamond&     1.66&     1.32&     1.01&     1.48&     1.96& &
			0.536&    0.420&    0.241&    0.153&    0.615\\
			GaAs&     1.69&     1.71&     1.57&     1.61&     4.09& &
			0.157&    0.206&    0.128&    0.124&    1.297\\
			Ge&     1.60&     1.71&     1.75&     1.69&     4.78& &
			0.113&    0.207&    0.135&    0.134&    1.657\\
			NaCl&     2.14&     0.78&     0.59&     0.70&     1.20& &
			0.438&    0.222&    0.170&    0.124&    0.237\\
			BaTiO${}_3$&     3.07&     2.68&     1.59&     1.76&     3.57& &
			2.742&    2.983&    1.992&    2.021&    2.705\\
			SrTiO${}_3$&     3.10&     2.67&     1.54&     1.79&     3.28& &
			3.017&    3.313&    2.151&    2.139&    2.466\\
			MnO&     2.16&     2.70&     1.78& \multicolumn{1}{c}{-}& \multicolumn{1}{c}{-}& &
			0.978&    1.295&    1.499& \multicolumn{1}{c}{-}& \multicolumn{1}{c}{-}\\
			\hline
			Mean&	2.23& 1.93& 1.42& 1.53& 3.10& &
			1.077&  1.164& 0.850& 0.728& 1.394
		\end{tabular}
	\end{ruledtabular}
	\label{table:iae_and_HK_metric}
\end{table*}
\subsection{Metrics for assessing densities and KS potentials}
In order to gain more quantitative insight into the quality of these densities and potentials, we now define two metrics to quantify the ``error'' between the QMC density/potential obtained via inversion and that of each DFA\@.
A simple comparison of densities can be carried out using the integrated absolute error (IAE) per electron,defined in similar fashion to the IAD in Eq.\ \eqref{eq:IAD},
\begin{equation}
	\textrm{IAE} = \frac{1}{N_\text{r}}\sum_{n=1}^{N_\text{r}}|\rho_\mathrm{QMC}(\vec{r}_n)-\rho_\mathrm{DFA}(\vec{r}_n)|.
	\label{eq:IAE}
\end{equation}
Additionally, given the one-to-one correspondence between the density and the Kohn-Sham potential, we define a second metric to compare both a pair of densities and their corresponding Kohn-Sham potentials simultaneously:
\begin{equation}
	\begin{split}
		\mathcal{E}
		=
		-\sum_\sigma\int\intd{r} \,
		\big( \rho^\sigma_\mathrm{DFA}(\vec{r}) - \rho^\sigma_\mathrm{QMC}(\vec{r}) \big)
		\\
		\times \big(v^\sigma_\mathrm{DFA}(\vec{r}) - v^\sigma_\mathrm{QMC}(\vec{r}) \big) > 0,
	\end{split}
	\label{eq:HK_metric}
\end{equation}
which for a numerical density stored on a grid of size $N_\text{r}$ is evaluated via
\begin{equation}
	\begin{split}
		\mathcal{E}
		=
		-\sum_\sigma\frac{1}{N_\text{r}}\sum_{n=1}^{N_\text{r}} \big(\rho^\sigma_\mathrm{DFA}(\vec{r}_n) - \rho^\sigma_\mathrm{QMC}(\vec{r}_n) \big)
		\\
		\times \big(v^\sigma_\mathrm{DFA}(\vec{r}_n) - v^\sigma_\mathrm{QMC}(\vec{r}_n)\big).
	\end{split}
\end{equation}
The use of the quantity $\mathcal{E}$ (with units of energy) is motivated by the proof
of the HK theorem, see e.g.\ Theophilou's Ref.~\onlinecite{Theophilou_1979_EDFT_HK},
and has the physical interpretation that a region within which the potential is deeper
must have a correspondingly higher density.
An advantage of this metric over a simple integrated difference in potentials similar to the IAE in Eq.\ \eqref{eq:IAE} is the appropriate weighting of the potential; in particular this ensures the oscillations in the QMC potential observed in the low density regions in some systems do not significantly contribute to the overall value of the metric.
As discussed in Ref.\ \onlinecite{Gidopoulos_SDFT:2007}, such oscillations in the KS potential do not pose a significant problem given that there is no charge density to `see' these oscillations.
In addition, we note that this $\mathcal{E}$ is invariant under constant shifts in the potential(s), or equivalently an overall constant shift in the potential difference $V_0$
\begin{align*}
	\mathcal{\tilde{E}} = &-\sum_\sigma \int\intd{r}~ \Delta \rho^\sigma(\vec{r})\big(\Delta v^\sigma(\vec{r}) + V_0\big)
	\\
	= &-\sum_\sigma \bigg[\int \Delta \rho^\sigma(\vec{r}) \Delta v^\sigma(\vec{r})~\intd{r}
	\addtocounter{equation}{1}\tag{\theequation}
	\label{eq:invariance_pot_diff}
	\\
	&+
	V_0\int \Delta \rho^\sigma(\vec{r})~\intd{r}
	\bigg]
	= \mathcal{E},
\end{align*}
where
$\Delta\rho^\sigma(\vec{r}) = \rho^\sigma_\mathrm{DFA}(\vec{r}) - \rho^\sigma_\mathrm{QMC}(\vec{r})$ and \\
$\Delta v^\sigma(\vec{r})=v^\sigma_\mathrm{DFA}(\vec{r}) - v^\sigma_\mathrm{QMC}(\vec{r})$.
Here, we make use of the fact that the integral on the last line of  Eq.~\eqref{eq:invariance_pot_diff} vanishes since each density integrates to $N$ electrons, thus $\int \Delta \rho(\vec{r})~\intd{r}=0$, hence Eq.~\eqref{eq:invariance_pot_diff} reduces to the original definition in Eq.~\eqref{eq:HK_metric}.

We have compared the densities from QMC with various DFAs using both the IAE and $\mathcal{E}$ in Table \ref{table:iae_and_HK_metric}. In order to facilitate a like-for-like comparison between systems, we have quoted the value for each metric per electron.
No finite size correction was applied to either the charge density or the potential obtained from inversion; instead we used the density obtained from the QMC calculation performed with the largest computational supercell.

Considering the IAE first, as one might expect the performance of the LDA ranges from being comparable in accuracy to other functionals from higher rungs of Jacob's ladder to being the worst performing functional, particularly in Si and NaCl.
At the other end of the spectrum, HF has, by some margin, the largest error out of all the DFAs studied. As seen in our results, this can be attributed to errors which arise from failing to account for correlation, and thus an overlocalization with respect to the QMC (or more generally exact) density.
The functional with the most accurate density according to this metric (ground state density being closest to the QMC density) across the widest range of systems is rSCAN, performing better than the HSE06 functional, despite its lower computational cost.
Then PBE, followed by LDA, with the worst functional in this metric being HF\@.

Turning our attention now to the energy metric $\mathcal{E}$, one might expect that the error in the density would be correlated with the error in the potential, resulting in similar trends to those observed in the IAEs\@.
However, this turns out not to be the case, even in systems with similar chemical environments.

For instance, according to the $\mathcal{E}$ metric,
in $\textrm{BaTiO}_3$ and $\textrm{SrTiO}_3$, HF and then LDA are superior to PBE,
but still worse than rSCAN and HSE06.
Furthermore, in some systems like MnO, rSCAN appears to have a worse potential despite having a more accurate density as evidenced by the larger value of $\mathcal{E}$ despite the lower IAE\@.
Overall, according to metric $\mathcal{E}$, the best functionals are rSCAN together with HSE06, followed by LDA, PBE, and finally HF\@.

\begin{table*}[ht]
	\begin{ruledtabular}
    \caption{
		DE and FE contributions to the total energy [see Eqs.\ \eqref{eq:func_err_defn} and \eqref{eq:density_err}] of density functional approximations (DFAs) compared to QMC results. All energies are quoted in millihartrees (mHa) with errors indicated in round brackets.
		For non-local DFAs, the GKS orbitals were used to evaluate $E_\mathrm{DFA}[\rho_\mathrm{DFA}] = E_\mathrm{DFA}[\{ \phi_{i}^\mathrm{GKS} [\rho]\}]$.
	}
		\begin{tabular}{ld{2}d{2}d{2}d{2}d{2}d{2}d{2}d{2}d{2}d{2}}
			\multirow{2}{*}{Material}&
			\multicolumn{2}{c}{LDA}& \multicolumn{2}{c}{PBE}&
			\multicolumn{2}{c}{rSCAN}& \multicolumn{2}{c}{HSE06}&
			\multicolumn{2}{c}{HF}\\
			&
			\multicolumn{1}{c}{FE}& \multicolumn{1}{c}{DE}&
			\multicolumn{1}{c}{FE}& \multicolumn{1}{c}{DE}&
			\multicolumn{1}{c}{FE}& \multicolumn{1}{c}{DE}&
			\multicolumn{1}{c}{FE}& \multicolumn{1}{c}{DE}&
			\multicolumn{1}{c}{FE}& \multicolumn{1}{c}{DE}
			\\ \hline
			Si&	37.40(2)&	-3.24(2)&	26.10(2)&	-3.03(2)&	13.91(2)&	-2.53(2)&	30.68(2)&	-8.76(2)&	309.58(2)&	-12.72(2)\\
			Diamond&	20.85(3)&	-2.79(3)&	0.94(3)&	-2.17(3)&	7.19(3)&	-2.32(3)&	58.95(3)&	-59.12(3)&	352.00(3)&	-11.36(3)\\
			GaAs&	-24.46(3)&	-1.20(3)&	-40.29(3)&	-1.23(3)&	-52.51(3)&	-1.11(3)&	-30.60(3)&	-9.34(3)&	245.32(3)&	-13.92(3)\\
			Ge&	-29.62(4)&	-0.86(4)&	-42.74(4)&	-1.21(4)&	-51.61(4)&	-1.16(4)&	-31.26(4)&	-9.53(4)&	253.52(4)&	-15.70(4)\\
			NaCl&	89.71(2)&	-2.58(2)&	37.60(2)&	-0.91(2)&	5.44(2)&	-0.94(2)&	37.23(2)&	-7.74(2)&	275.28(2)&	-3.60(2)\\
			BaTiO\textsubscript{3}&	65.1(3)&	-46.5(3)&	-137.0(3)&	-48.0(3)&	-107.0(3)&	-29.5(3)&	-56.4(3)&	-39.8(3)&	1035.9(3)&	-62.4(3)\\
			SrTiO\textsubscript{3}&	95.6(3)&	-51.3(3)&	-100.7(3)&	-52.5(3)&	-73.9(3)&	-32.0(3)&	-21.3(3)&	-42.9(3)&	1066.6(3)&	-56.2(3)\\
			MnO& -1138.9(3)&	-16.4(3)&	-1430.1(3)&	-20.6(3)&	-802.2(3)&	-12.6(3)&
			\multicolumn{1}{c}{-}&\multicolumn{1}{c}{-}&
			\multicolumn{1}{c}{-}&\multicolumn{1}{c}{-}\\
		\end{tabular}
	\end{ruledtabular}
	\label{table:de_fe_values}
\end{table*}

\subsection{Density-functional error analysis}
In addition to the density and the potential, we can also assess the error in the total energy as a result of the approximation for $\exc$ in order to gain further insight into the successes (or failures) of DFAs.
We follow the error analysis scheme of Burke and coworkers \cite{Burke_DE_FE:2013,Burke_DE_FE:2014,Burke_DE_FE:2017,Burke_DE_FE:2018,Burke_DE_FE:2019}
for the total energy error of a particular DFA,
\begin{equation}
    	\Delta E_\mathrm{DFA}
        \doteq
        E_\mathrm{DFA}[\rho_\mathrm{DFA}] - E_\mathrm{exact}[\rho_\mathrm{exact}] ,
\end{equation}
where $\rho_\mathrm{exact}(\vec{r})$ is the exact density and $\rho_\mathrm{DFA}(\vec{r})$ is the density obtained via a self-consistent calculation using the DFA; $E_\mathrm{DFA}[\rho_i]$ is the DFA total energy evaluated at the density $\rho_i(\vec{r})$ and $E_\mathrm{exact}[\rho_\mathrm{exact}]$ is the exact total energy.
Next, the total energy error is partitioned into two separate errors due to (i) an inaccurate density [density-driven error (DE)] and (ii) an inaccurate XC energy functional [functional error (FE)],
\begin{equation}
	\Delta E_\mathrm{DFA} = \Delta E_\mathrm{FE} + \Delta E_\mathrm{DE} ,
\end{equation}
with
\begin{align}
	\Delta E_\mathrm{FE} &= E_\mathrm{DFA}[\rho_\mathrm{exact}] -  E_\mathrm{exact}[\rho_\mathrm{exact}],
	\label{eq:func_err_defn}
	\\
	\Delta E_\mathrm{DE} &= E_\mathrm{DFA}[\rho_\mathrm{DFA}] -  E_\mathrm{DFA}[\rho_\mathrm{exact}]<0. \label{eq:density_err}
\end{align}
$\Delta E_\mathrm{DE}<0$ because $\rho_\mathrm{DFA}(\vec{r})$ is the minimizing density of the functional $E_\mathrm{DFA}[\rho]$.
In our analysis, the exact density $\rho_\mathrm{exact}(\vec{r})$ is approximated by the QMC density $\rho_{\text{QMC}}(\vec{r})$, while the exact energy $E_\mathrm{exact}[\rho_\mathrm{exact}]$ is taken to be the DMC energy extrapolated to zero time step and infinite system size. We extrapolate the DMC energy linearly to zero time step \cite{Needs:20}.
The zero-time step extrapolated values for each supercell are then extrapolated to infinite system size using Eq.\ \eqref{eq:lin_fit}.

To proceed, we distinguish between local/semilocal approximations (LDA/PBE) and nonlocal approximations (rSCAN/HSE06/HF)\@. For the former, the application of Eqs.~(\ref{eq:func_err_defn}) and (\ref{eq:density_err}) is straightforward, but not for the latter.
For example, in nonlocal approximations $E_\mathrm{DFA}[\rho_\mathrm{exact}]$ must be obtained from an inversion of the QMC density with
the appropriate nonlocal GKS equations, using the GKS orbitals,
$E_\mathrm{DFA}[\rho_\mathrm{exact}] = E_\mathrm{DFA}[\{ \phi_{i}^\mathrm{GKS} [\rho_\mathrm{exact}] \}]$.
However, in this work, we only perform a standard KS inversion of the QMC density with a local KS potential and obtain KS orbitals $\{ \phi_{i}^\mathrm{KS} [\rho_\mathrm{exact}] \}$.
To estimate the FE [Eq.\ \eqref{eq:func_err_defn}] in terms of the calculated quantities we write
\begin{multline}
        \Delta E_\mathrm{FE} = \Delta E_\mathrm{DFA} ^\mathrm{NL} [\rho_\mathrm{exact}] \\
     + E_\mathrm{DFA}[\{ \phi_{i}^\mathrm{KS} [\rho_\mathrm{exact}] \}] -
    E_\mathrm{exact}[\rho_\mathrm{exact}] ,
 \end{multline}
where
\begin{equation}
    \Delta E_\mathrm{DFA} ^\mathrm{NL} [\rho] \doteq
    E_\mathrm{DFA}[\{ \phi_{i}^\mathrm{GKS} [\rho]\}] -
    E_\mathrm{DFA}[\{ \phi_{i}^\mathrm{KS} [\rho] \}]\leq 0
\end{equation}
is the DFA's nonlocality energy at density $\rho(\vec{r})$ (see Ref.~\onlinecite{Ravindran:2024}).
The KS orbitals $\{ \phi_{i}^\mathrm{KS} [\rho] \}$ are found by inverting the density $\rho$ with a local potential.
Then, to obtain the FE, we approximate
\begin{equation} \label{nl_approx}
    \Delta E_\mathrm{DFA} ^\mathrm{NL} [\rho_\mathrm{exact}] \simeq
    \Delta E_\mathrm{DFA} ^\mathrm{NL} [\rho_\mathrm{DFA}] ,
\end{equation}
and we conclude
\begin{multline}
        \Delta E_\mathrm{FE} \simeq \Delta E_\mathrm{DFA} ^\mathrm{NL} [\rho_\mathrm{DFA}] \\
     + E_\mathrm{DFA}[\{ \phi_{i}^\mathrm{KS} [\rho_\mathrm{exact}] \}] -
    E_\mathrm{exact}[\rho_\mathrm{exact}] .
 \end{multline}
We note that the nonlocality energy vanishes for local/semilocal DFAs (LDA/PBE)\@.
In general $\Delta E_\mathrm{DFA} ^\mathrm{NL} [\rho] \le 0$, because the minimum energy
of a nonlocal DFA is attained for the GKS orbitals.

Similarly, the DE [Eq.~\eqref{eq:density_err}] can be written as
 \begin{multline}
        \Delta E_\mathrm{DE} =
        \Delta E_\mathrm{DFA} ^\mathrm{NL} [\rho_\mathrm{DFA}]
        - \Delta E_\mathrm{DFA} ^\mathrm{NL} [\rho_\mathrm{exact}]
        \\
        +E_\mathrm{DFA}[\{ \phi_i^\mathrm{KS} [\rho_\mathrm{DFA}] \}]
        -E_\mathrm{DFA}[\{ \phi_i^\mathrm{KS} [\rho_\mathrm{exact}] \}].
 \end{multline}
Using Eq.~\eqref{nl_approx} we have
\begin{equation}
        \Delta E_\mathrm{DE} \simeq
                E_\mathrm{DFA}[\{ \phi_{i}^\mathrm{KS} [\rho_\mathrm{DFA}] \}]
        -E_\mathrm{DFA}[\{ \phi_{i}^\mathrm{KS} [\rho_\mathrm{exact}] \}].
\end{equation}

The calculated FEs and DEs for the DFAs considered in this work are given in Table \ref{table:de_fe_values}. For most of the systems we studied, $\Delta E_\mathrm{FE}>0$, except in GaAs, Ge, and MnO\@.
Unlike the DE, there is no strict inequality for the FE [cf.\ Eqs.~\eqref{eq:func_err_defn} and \eqref{eq:density_err}].

With the exception of HF, the sign of the FE for a given system was found to be
the same for all DFAs in this work, possibly because these DFAs are typically constructed from the LDA via a so-called enhancement factor \cite{PBE,HSE,SCAN,RSCAN} and reduce to the LDA in the appropriate limits.
A more thorough study across a wider range of systems and DFAs is required to see if this trend is systematic.

In all the systems in our study, the DE was smaller in magnitude than the FE, in some cases by one to two orders of magnitude, with the notable exception of PBE for
diamond where the DE was larger than the FE\@.
Moreover, we found that rSCAN gave the lowest value for $\Delta E_\mathrm{DE}$ across the systems studied, with the exception of Ge, where LDA had the lowest value. This is consistent with our findings that rSCAN had the smallest IAE and joint lowest $\mathcal{E}$\@.
By contrast, the FE exhibited more variation; in the perovskites BaTiO$_3$ and SrTiO$_3$, HSE06 had the lowest functional error followed by rSCAN in BaTiO$_3$
but LDA in SrTiO$_3$.
Outside these two perovskites, rSCAN was also the DFA with the lowest FE,
except in Ge and GaAs, where LDA had the lowest FE\@.
It is however interesting to note that despite the larger FEs and DEs,
HSE06 often has lower total errors compared to rSCAN, notably in diamond, due to cancellation of DEs and FEs.

\begin{table*}[t!]
	\caption{
		GKS and KS gaps (in eV) for various DFAs. The KS gaps are obtained via inversion of the GKS density.
		The KS values for QMC are extrapolated to infinite system size (see Table\ \ref{table:supercell_gaps}).
		In addition to the (`exact') KS gap obtained from inversion of the QMC density, the QMC fundamental gap $E_\mathrm{g}^\mathrm{QMC}$ is listed where available.
		Mean absolute errors (MAEs) are quoted for each DFA relative to the QMC KS gap.
		Experimental (Expt.)\ values are taken from Ref.\ \onlinecite{Madelung:1996} unless otherwise stated.
	}
	\begin{ruledtabular}
		\begin{tabular}{ld{2}d{2}d{2}d{2}d{2}d{2}d{2}d{2}d{2}d{2}d{2}}
			{}&
			\multicolumn{1}{c}{LDA}& \multicolumn{1}{c}{PBE}& \multicolumn{2}{c}{rSCAN}& \multicolumn{2}{c}{HSE06}& \multicolumn{2}{c}{HF}&
			\multicolumn{2}{c}{Quantum Monte Carlo}&
			\multicolumn{1}{c}{Expt.}
			\\
			{}&
			\multicolumn{1}{c}{KS}& \multicolumn{1}{c}{KS}&
			\multicolumn{1}{c}{KS}& \multicolumn{1}{c}{GKS}&
			\multicolumn{1}{c}{KS}& \multicolumn{1}{c}{GKS}&
			\multicolumn{1}{c}{KS}& \multicolumn{1}{c}{GKS}&
			\multicolumn{1}{c}{KS}& \multicolumn{1}{c}{$E_\mathrm{g}^\mathrm{QMC}$}&
			{}
			\\ \hline
			Si & 0.46 & 0.63 & 0.70 & 0.76 & 0.70 & 1.13 & 1.20 & 6.17 &
			0.81& \multicolumn{1}{c}{1.8(1)\footnotemark[1],
				1.50(2)\footnotemark[2]
			}
			& 1.17\\
			Diamond & 3.99 & 4.22 & 4.30 & 4.31 & 4.28 & 5.34 & 4.74 & 12.42&
			4.41 & \multicolumn{1}{c}{6.8(1)\footnotemark[1],5.94(4)\footnotemark[3],5.73\footnotemark[4]}
			& 5.5\\
			GaAs & 1.16 & 1.38 & 1.46 & 1.60 & 1.46 & 1.94 & 1.96 & 7.07&
			1.37 & &{} 1.52\\
			Ge & 0.26 & 0.43 & 0.49 & 0.56 & 0.51 & 0.88 & 1.03 & 5.73 &
			0.38 & {}& 0.79\\
			NaCl & 4.94 & 5.36 & 5.50 & 6.00 & 5.61 & 6.68 & 6.15 & 13.76  &
			5.67& {}& 8.97\\
			$\textrm{BaTiO}_3$ & 1.81 & 2.24 & 2.56 & 2.66 & 2.78 & 3.46 & 4.77 & 11.73 &
			3.21 &  {}& 3.2\footnotemark[5]\\
			$\textrm{SrTiO}_3$ & 1.89 & 2.35 & 2.69 & 2.80 & 2.90 & 3.49 & 4.87 & 11.72 &
			3.42& {}& 3.25\footnotemark[6]\\
			MnO & 2.89 & 3.23 & 3.24 & 3.32 &
			\multicolumn{1}{c}{-}& \multicolumn{1}{c}{-}&
			\multicolumn{1}{c}{-}& \multicolumn{1}{c}{-}&
			2.95 &
			\multicolumn{1}{c}{4.8(2)\footnotemark[7]}
			& 3.9\footnotemark[8]
			\\
		\hline
		\text{MAE}& 0.60& 0.39& \multicolumn{2}{c}{0.29}& \multicolumn{2}{c}{0.22}& \multicolumn{2}{c}{0.78}\\
		\end{tabular}
	\end{ruledtabular}
	\footnotetext[1]{Using Slater-Jastrow (SJ) result from Ref.\ \onlinecite{Ceperley_Silicon_Diamond:2020}}
	\footnotetext[2]{Ref.\ \onlinecite{Annaberdiyev:2021}}
	\footnotetext[3]{Ref.\ \onlinecite{Ceperley_Diamond_Gap:2023}}
	\footnotetext[4]{Ref.\ \onlinecite{Towler_Diamond_Gap:1999}}
	\footnotetext[5]{Ref.\ \onlinecite{BaTiO3_band_gap}}
	\footnotetext[6]{Ref.\ \onlinecite{SrTiO3_band_gap}}
	\footnotetext[7]{Ref.\ \onlinecite{MnO_QMC_gap:2004} using Ne core pseudopotentials}
	\footnotetext[8]{Ref.\ \onlinecite{MnO_band_gap}}
	\label{table:dfa_gaps}
\end{table*}

Finally it comes as no surprise that HF is the worst overall with regard to FE as the lack of correlation energy in the HF functional leads to FEs that are one to two orders of magnitude larger compared to standard DFAs.
Nonetheless, it is noteworthy that in spite of this much larger FE, the DE for HF remains comparable to other DFAs.

\section{Band gaps and the exchange-correlation derivative discontinuity \label{section:discontinuity}}
It is well-known that the (exact) KS band gap $E_{\mathrm{g},s}$ is not equal to the fundamental gap $E_\mathrm{g}$ of the system, but rather
\begin{equation}
	E_\mathrm{g} = E_{\mathrm{g},s} + \Delta_{xc},
\end{equation}
where $\Delta_{xc}$ is the XC derivative-discontinuity \cite{PPLB_Delta_XC:1982,Perdew_Delta_XC:1997,GKS_OEP_SCAN_Perdew_2016,band_gaps_in_GKS_2017,Sham_Shulter:1983,Sham_Shulter:1985,Knorr_Godby:1992,Knorr_Godby:1994}.
The relative contributions of $\Delta_{xc}$ and $E_{\mathrm{g},s}$
to $E_\mathrm{g}$ in solids are generally unknown.
In particular, it is believed that in strongly-correlated systems dominated by Mott physics, $E_{\mathrm{g},s}$ is zero or negligible, such that the contribution to the fundamental gap $E_\mathrm{g}$ is largely from the exchange-correlation derivative discontinuity $\Delta_{xc}$.

Benchmarking the KS gap compared to other properties such as lattice parameters and cohesive energies is particularly challenging for similar reasons as the exact KS potential $\vks$, namely that it is not an experimental observable.
Although inversion offers a potential approach to gauge the size of $E_{\mathrm{g},s}$,
the gaps obtained are highly sensitive to the pseudopotentials used for the calculation, and particularly to the treatment of semicore states as (frozen) core states, as discussed in Appendix \ref{appendix:pspot_core_effects}.

In Table \ref{table:dfa_gaps}, we report both KS and GKS gaps for the systems considered in this work. Note that the KS gaps reported for the rSCAN, HSE, and HF functionals are obtained via inversion of the respective GKS target densities.

However, although XC potentials $\vxcs(\vec{r})$ of a nonlocal DFA, obtained via inversion of the DFA GKS density, are similar to the KS XC potentials obtained via OEP \cite{Ravindran:2024,Hollins:2017_LFX,Hollins:2017_LFX_metals}, the inverted XC potentials do not strictly have an XC discontinuity $\Delta_{xc}$,
because the inverted potentials are not functional derivatives of an XC energy functional $\exc[\rho]$. The same of course holds for the XC potentials obtained by inversion of the QMC density.

Nonetheless, based on the similarity of the inverted XC potentials with OEP for a nonlocal DFA, we will loosely refer to the XC discontinuity of the inverted XC potentials.

Firstly, we note that the KS gap obtained via inversion of the DFA GKS target density is always smaller than the GKS gap since the latter approximately incorporates $\Delta_{xc}$. The degree to which the GKS and KS gaps differ is a measure of the nonlocality of a given DFA within GKS, as discussed in greater detail in Ref.\ \onlinecite{Ravindran:2024}.

In most of the systems studied, the size of the KS gap tends to follow the progression of functionals in the
``Jacob's ladder of density functional approximations''
\cite{Jacobs_ladder}, with LDA having the smallest gap, followed by PBE (GGA), rSCAN (meta-GGA), and HSE06 (hybrid DFA)\@.
The KS gap obtained from inversion of the HF density is the largest, being larger than the reference KS gap from inversion of the QMC density.
This error is attributed to the lack of correlation.
Based on the systems studied, we further found that the mean absolute error of each DFA's KS gap relative to the KS-QMC gap likewise follows the ascendancy of Jacob's ladder, with KS-HSE06 having the lowest mean absolute error (MAE) of 0.22~eV and KS-rSCAN following closely behind with an MAE of 0.29~eV\@.

However, we note that the KS gap from QMC is not always larger than the KS gaps of DFAs (besides HF-KS),
particularly in MnO, where both the PBE gap and the KS-rSCAN gap are larger than the KS-QMC gap.
Transition metals are well-known within both DFT and QMC to be particularly sensitive to the treatment of the semicore states \cite{MnO_QMC_gap:2004,TMO_QMC_review,Mitas_Fe_atom_QMC:1994},
and this is no different for Mn with the treatment of the $3s$ and $3p$ states as frozen core states.

For completeness, we remark that the DFA KS gaps for MnO change significantly when using the
LDA pseudopotentials \footnote{These pseudopotentials are shipped with CASTEP as \texttt{MnO\_00.recpot} and \texttt{O\_00.recpot} for Mn and O, respectively.} of M.-H.\ Lee \cite{Lee_CASTEP_recpots_thesis:1995} instead of the TN Dirac-Fock pseudopotentials, despite both pseudopotentials treating the same electronic states as core states, differing only in the level of theory at which they are generated.
For instance, the LDA KS gap changes from 2.89~eV with the TN pseudopotentials to 1.12~eV with the LDA pseudopotentials, the PBE KS gap changes from 3.23~eV to 1.71~eV
and the rSCAN KS gap changes from 3.24~eV to 1.87~eV\@.

In the case of nonlocal DFAs, an estimate of the derivative discontinuity $\Delta_{xc}$ (or $\Delta_x$ in the case of HF) is
obtained from the difference between the GKS and KS values for each functional.
In particular, we can see that although HSE06 and rSCAN have comparable performance with regards to the accuracy of the KS gap (relative to KS-QMC),
the GKS-HSE06 gaps are larger than the GKS-rSCAN gaps.
This is due to %
the use of the nonlocal Fock exchange operator (see also Refs.\ \onlinecite{GKS_OEP_SCAN_Perdew_2016,band_gaps_in_GKS_2017}).
In Ref.~\onlinecite{Ravindran:2024}, it was found that the GKS-HSE06 had a lower MAE relative to experimental results for a
range of semiconductors and insulators
(a subset of which we study in our present work); based on the results in this study, the higher accuracy
of HSE06 appears to be due to the better estimate
of $\Delta_{xc}$ in GKS rather than an improvement in the density or the XC potential $\vxcs(\vec{r})$,
at least in the systems studied.

It is also worth noting that the differences in KS gaps arise almost entirely from differences in $\vxcs(\vec{r})$, since
the differences in $v_{H}(\vec{r})$ were found to be relatively small.

The magnitude of the full $\Delta_{xc}$ itself could in principle be inferred by comparing the KS-QMC gaps to experimental gaps.
However, care should be taken when doing this due to pseudopotential effects as well as physical effects such as band gap renormalization
due to electron-phonon interactions and phonon zero-point renormalization that are ignored in our calculations, especially for diamond \cite{Giustino_2010_Diamond_EPI,Ceperley_Silicon_Diamond:2020,Kresse_ZPR_Semiconductors:2018}.
An alternative approach is to compare with results from more accurate electronic structure methods, but this can present its own problems.
For example, in QMC the presence of finite-size errors can amount to a considerable systematic error \cite{Annaberdiyev:2021} and the treatment of such errors is crucial for the accurate calculation of gaps as shown by the spread of QMC (fundamental) gaps $E_\mathrm{g}^\mathrm{QMC}$ in Table \ref{table:dfa_gaps}.
The high cost of such calculations has likewise limited their use, with only a few QMC gaps being reported in the literature.

\section{Conclusions \label{section:conclusions}}
We have performed density inversion of QMC densities and obtained accurate KS potentials $\vks(\vec{r})$ and, by extension, accurate XC potentials $\vxcs(\vec{r})$ for several insulating solids given the pseudopotentials used in the calculations.
The accuracy of the inverted KS potentials depends on the quality of the target QMC densities, including the accurate treatment of finite-size errors as well as uncontrolled errors resulting from the treatment of semicore states as frozen core states in the pseudopotentials.

The QMC densities and the KS potentials obtained therefrom via inversion serve as benchmarks with which to compare popular DFAs at each rung of Jacob's ladder of functionals, namely the LDA, PBE (GGA), rSCAN (meta-GGA), HSE06 (hybrid DFA), as well as the HF approximation.
For comparison, we have used two metrics, the IAE as defined in Eq.\ \eqref{eq:IAE} and the energy $\mathcal{E}$ as defined in Eq.\ \eqref{eq:HK_metric}, in which the potential difference is weighted by the density difference.

For the systems we studied, we found that although the DFAs produce similar electron densities, the corresponding differences in the exchange-correlation potential $\vxcs(\vec{r})$ can still be substantial.
This is especially evident in the perovskite systems $\textrm{BaTiO}_{3}$ and $\textrm{SrTiO}_{3}$, where regions with only slightly higher QMC densities compared to the DFAs gave rise to deeper and more attractive $\vxcs(\vec{r})$ potentials.

Furthermore, we find that the differences among the KS potentials $\vks(\vec{r})$ of the same system calculated using different methods stem primarily from variations in $\vxcs(\vec{r})$, while the differences in the Hartree potential $ v_H(\vec{r})$ are comparatively minor due to the overall similarity in electron densities across these methods.
As a result, the discrepancies in the KS gap $ E_{\mathrm{g},s}$, and likely in other properties as well, can be attributed specifically to changes in $\vxcs(\vec{r})$.
This occurs despite the fact that the overall variation in $ \vks(\vec{r}) $ remains small, since $\vxcs(\vec{r})$, while relatively minor in magnitude, has a clearly significant contribution.

As expected, HF yields the highest IAE per electron among all DFAs, due to its tendency to overlocalize the electronic density as a result of neglecting electron correlation.
However, according to the second metric $\mathcal{E}$ defined in Eq.\ \eqref{eq:HK_metric}, the LFX potential obtained from the inversion of the HF density turns out to perform reasonably well compared with other DFAs, even though it is still an exchange-only potential.

In addition to comparing the potential and the density, we assess the quality of the approximate XC energy functional $\exc[\rho]$ of each DFA by following the density functional error analysis of Burke and coworkers \cite{Burke_DE_FE:2013,Burke_DE_FE:2017,Burke_DE_FE:2018,Burke_DE_FE:2019}, noting that our analysis distinguishes between standard and generalized KS schemes.
In a similar vein to the previous observations, we find that the performance of functionals with regards to the DE $\Delta E_\mathrm{DE}$ and the FE $\Delta E_\mathrm{FE}$ is not uniform; that is to say, functionals that have a lower FE $\Delta E_\mathrm{FE}$ do not necessarily have a lower DE $\Delta E_\mathrm{DE}$ as well.

We have also calculated the value of the KS gaps $E_{\mathrm{g},s}^\mathrm{DFA}$ for each DFA (as obtained via density inversion for nonlocal DFAs), and compared them against the corresponding KS gap $E_{\mathrm{g},s}$ from the inversion of the QMC density, which serves as an approximation to the ``exact'' $E_{\mathrm{g},s}$.
This comparison provides another metric to assess the quality of the DFAs. For the rSCAN and HSE06 functionals the KS gaps are similar to the KS-QMC gaps.

Finally, the improved accuracy of rSCAN and HSE06 in the calculation of the fundamental gap $E_\mathrm{g}$ is due to the approximate
incorporation of the XC derivative discontinuity $\Delta_{xc}$ within the GKS calculation.
The difference between the KS and GKS gaps for a given DFA is an estimate of the exchange and correlation derivative discontinuity $\Delta_{xc}^\text{DFA}$ for that DFA\@.
In particular for HF the difference between the LFX (KS-HF) gap and the HF gap gives the exchange discontinuity $\Delta_{x}$.

In principle the KS-QMC gaps could be compared to experimental gaps in order to infer the size of the XC discontinuity $\Delta_{xc}$  of the exact $\vxcs(\vec{r})$ for these systems.
However, this approach is prone to introducing systematic errors from the use of pseudopotentials in the KS-QMC gap and the neglect of other physical effects present in the fundamental gap $E_\mathrm{g}$, such as the phonon corrections (zero-point renormalization, thermal phonon renormalization, electron-phonon coupling), which are known to be significant e.g.\ in diamond.

\begin{acknowledgments}
    We are grateful to Samuel P.\ Ladd for fruitful discussions.
    V.\ Ravindran acknowledges the Durham Doctoral Scholarship program for financial support. We acknowledge the use of the Durham Hamilton HPC service (Hamilton) and the UK national supercomputing facility (ARCHER2) funded by EPSRC grant EP/X035891/1. Further resources were provided by Lancaster University's high-end computing cluster.
\end{acknowledgments}

\appendix

\section{Effect of treatment of semicore states in calculated KS gaps \label{appendix:pspot_core_effects}}

For consistency, the same set of pseudopotentials is used for all calculations on a given system, fixing the external potential term in the KS Hamiltonian. Consequently, when comparing KS (XC) potentials obtained from the inversion of QMC densities with those from other DFAs, any differences arise from the functionals themselves rather than from variations in pseudopotentials, ensuring that each (pseudo-)material is treated on an equal footing within each formalism.

However, the calculation of the KS gap can be highly sensitive to how semicore states are treated within the pseudopotential formalism, in particular whether they are treated as part of the (frozen) core or as valence.
Since the TN pseudopotentials are obtained at the HF level of theory, to gauge the effect of the treatment of semicore states within the pseudopotential, we compared
LFX band structures using the TN pseudopotential to those using \textsc{castep} NCP19 on-the-fly-generated (OTFG) pseudopotentials (see Ref.\ \onlinecite{new_castep_pspots:2016}).
The character of each band was determined via partial density of states  (PDOS) calculations performed using the OptaDOS code \cite{optados}, which implements the population analysis methodology of Segall \textit{et al.}\ \cite{Segall_Population_Analysis}.
The adaptive broadening scheme of Yates \textit{et al.}\ \cite{Yates_Optados_Adaptive_Smearing} was applied within the calculation of the PDOS\@.
The resulting LFX band gaps are given in Table \ref{table:pspot_compare} along with the valence electronic configuration for the pseudopotentials.

\begin{table}[htbp!]
  \caption{
   LFX (KS-HF) band gaps calculated from the inversion of the HF density using \textsc{castep} NCP19 pseudopotentials and \textsc{casino} TN pseudopotentials along with the valence electronic configurations used in the pseudopotentials.
   An underline indicates states that are treated as valence using NCP19 pseudopotentials but as part of a frozen core in the TN pseudopotentials.
  }
 \begin{ruledtabular}
  \begin{tabular}{ld{2}d{2}c}
   Material &\multicolumn{1}{c}{TN\footnote{This work}} &\multicolumn{1}{c}{NCP19\footnote{Ref.\ \onlinecite{Ravindran:2024}}} & \begin{tabular}[c]{@{}c@{}}Valence\\ configuration\end{tabular}
   \\
   \hline
   Si & 1.20 & 1.17 & Si - $3s^2\;3p^2$\\
   Diamond & 4.74 & 4.75 & C - $2s^2\;2p^2$\\
   GaAs\footnote{Changes from direct to indirect gap using TN pseudopotentials} & 1.96 & 0.93 &
   \begin{tabular}[c]{@{}c@{}}Ga - $\underline{3d^{10}}\;4s^2\;4p^1$\\
    As - $\underline{3d^{10}}\;4s^2\;4p^3$
   \end{tabular}
   \\
   Ge & 1.03 & 0.46 & Ge - $\underline{3d^{10}}\;4s^2\;4p^2$\\
   NaCl & 6.15 & 6.21 &
   \begin{tabular}[c]{@{}c@{}}
    Na - $\underline{2s^2\;2p^6}\;3s^1$
    \\
    Cl - $3s^2\;3p^5$
   \end{tabular} \\
   BaTiO\textsubscript{3}& 4.77 & 4.07 & \begin{tabular}[c]{@{}c@{}}
    Ba - $\underline{5s^2\;5p^6}\;6s^2$
    \\
    Ti - $\underline{3s^2\;3p^6}\;4s^2\;3d^2$
    \\
    O - $2s^2\;2p^4$
   \end{tabular} \\
   SrTiO\textsubscript{3}\footnote{Ti and O valence configuration same as BaTiO\textsubscript{3}.}
   & 4.87 & 4.25 & Sr - $\underline{4s^2\;4p^6}\;5s^2$
  \end{tabular}
  \label{table:pspot_compare}
 \end{ruledtabular}
\end{table}

\begin{figure}[htbp!]
 \centering
 \includegraphics[width=\columnwidth]{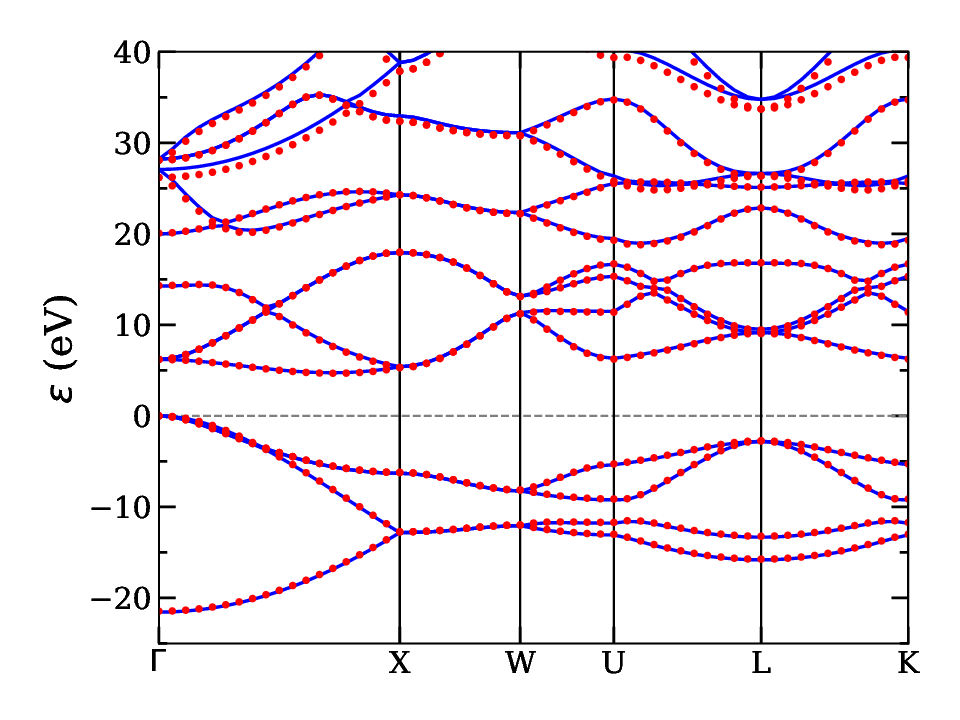}
 \caption{Calculated LFX band structure of diamond using TN (blue, solid line) and NCP19 (red, dotted line) pseudopotentials. Note that the Fermi energy has been set to 0 eV\@.
 }
 \label{fig:diamond_CASTEP_vs_CASINO_pspot}
\end{figure}

In both Si and diamond, where the valence electronic configurations in TN and NCP19 pseudopotentials are identical, there is a negligible difference between the band gaps and indeed the band structures as shown in Fig.\ \ref{fig:diamond_CASTEP_vs_CASINO_pspot} for diamond, although larger deviation is obtained for higher unoccupied states that lie far away from the Fermi energy.

\begin{figure}[htbp!]
 \centering
 \includegraphics[width=0.95\columnwidth]{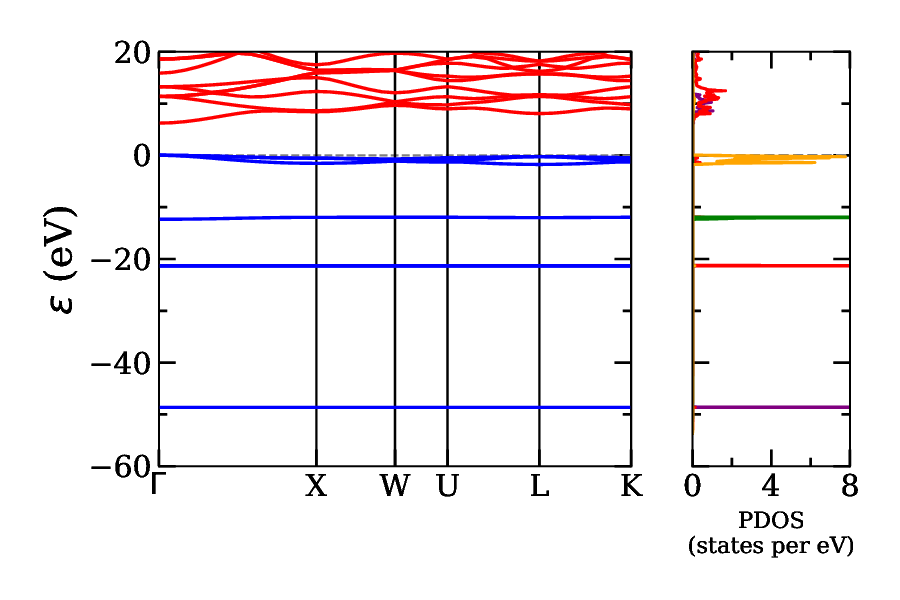}
 \caption{
  Calculated LFX band structure and PDOS of NaCl using NCP19 pseudopotentials. Note that the Fermi energy has been set to 0 eV\@.
  Left: blue indicates occupied bands and red unoccupied; right: purple Na $2s$, red Na $2p$, green Cl $3s$, and orange Cl $3p$ states.
  }
 \label{fig:NaCl_LFX_pdos}
\end{figure}

In NaCl, the Na $2s$ and $2p$ states can be considered as core states without significantly altering the results due to the large energy gap of over 10 eV between the lowest lying (Cl $3s$) state and the Fermi level. The Na $2s$ and $2p$ states lie even further below the Cl $3s$ states as shown in Fig.\ \ref{fig:NaCl_LFX_pdos}, with the flat bands demonstrating the strong localization of these states that one might expect.

\begin{figure*}[htbp!]
 \centering
 \subfloat[\label{subfig:GaAs_CASTEP_pspot}]{\includegraphics[width=0.95\columnwidth]{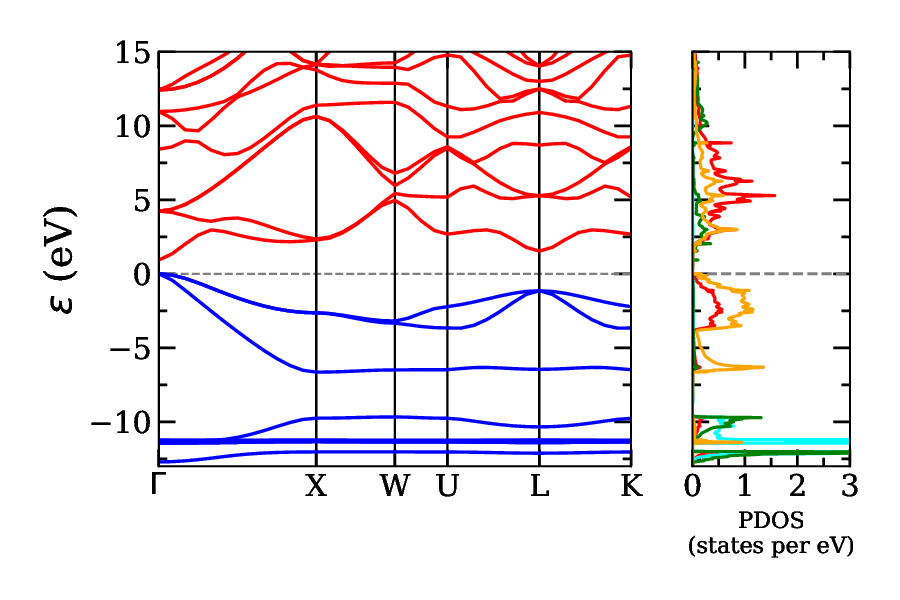}}
 \hspace{0.5em}
 \subfloat[\label{subfig:GaAs_CASINO_pspot}]{\includegraphics[width=0.95\columnwidth]{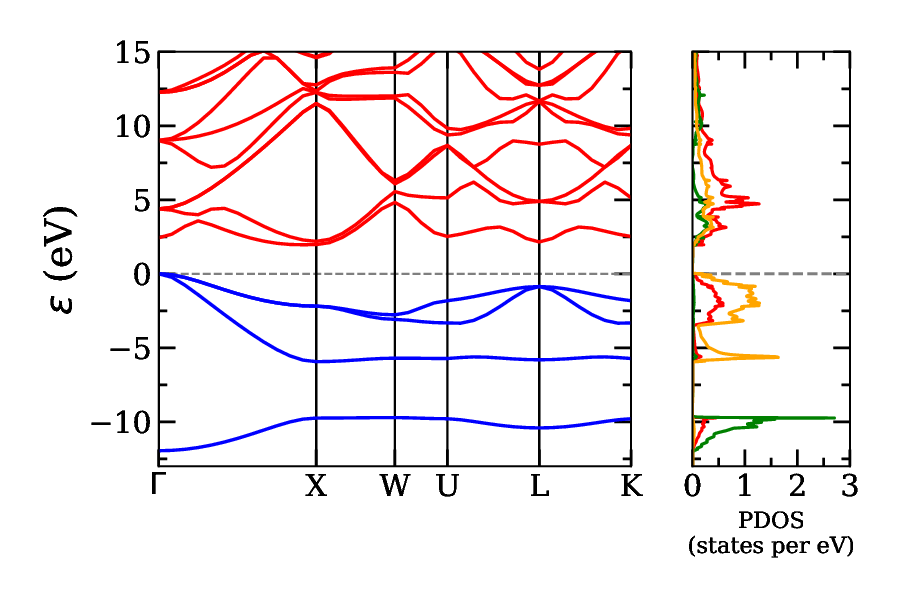}}
 \caption{
  Calculated LFX band structure and PDOS of GaAs using (a) NCP19 pseudopotentials and (b) TN pseudopotentials. Note that the Fermi energy has been set to 0 eV in each band structure.
  The color scheme for the band structure (left) is the same as that of  Fig.\ \ref{fig:NaCl_LFX_pdos}. The color scheme for the PDOS (right) is as follows. Red: Ga $4p$; cyan: Ga $3d$; green: As $4s$; and orange: As $4p$.
  Note that the LFX gap changes from a direct gap at $\Gamma$ with NCP19 pseudopotentials to an indirect gap at $\Gamma \to X$ with TN pseudopotentials.
 }
 \label{fig:GaAs_CASTEP_vs_CASINO_pspot}
\end{figure*}

On the other hand, in GaAs, the Ga $3d$ states turn out to be semicore states and consequently their treatment in a frozen core can affect the calculated electronic structure, and in particular the calculated KS gap.
This can be seen in both the band structures and PDOS in Fig.\ \ref{fig:GaAs_CASTEP_vs_CASINO_pspot} for LFX calculations, where the $3d$ states are treated as valence by NCP19 pseudopotentials in Fig.\ \ref{subfig:GaAs_CASTEP_pspot}
but as frozen core states by the TN pseudopotentials in Fig.\ \ref{subfig:GaAs_CASINO_pspot}.
The LFX gap not only changes in value but also changes from direct to indirect, with a direct gap of $0.93$ eV at the $\Gamma$ point using NCP19 pseudopotentials becoming an indirect gap of $1.96$ eV from $\Gamma \to X$ using TN pseudopotentials.
The direct gap for the TN pseudopotentials is $2.43$ eV\@.
In addition to the gap, the actual dispersion of the individual bands is also altered.
We find that for NCP19 and TN pseudopotentials the valence bandwidths are $3.19$ eV and $2.77$ eV, respectively, and the conduction bandwidths are $4.03$ eV and $2.86$ eV, respectively.

These results highlight the importance of treating accurately the semicore states, particularly, when comparing with experimental results.
We remind the reader that in our work, we use the same pseudopotential and lattice parameters across DFA and QMC calculations to maintain a consistent external potential.
This allows us to draw meaningful conclusions from the comparison between the DFA and the benchmark QMC results.

\section{Nonlinear core corrections for the exchange-correlation potential}\label{appendix:nlcc_effects}
Within the pseudopotential formalism, the total electronic density $\rho_\mathrm{total}(\vec{r})$ is partitioned into the core $\rho_\mathrm{core}(\vec{r})$ and the valence $\rho_\mathrm{val}(\vec{r})$ densities, the former of which is typically frozen in calculations.
The pseudopotential is constructed by first performing an all-electron atomic calculation to obtain the `screened' potential $v_l^\mathrm{total}(\vec{r})$ that is seen by the valence charge density in the atom $\rho_\mathrm{val}^\mathrm{atom}(\vec{r})$.
The pseudopotential $v_l^\mathrm{ion}(\vec{r})$ for the bare ion associated with the nuclear charge plus the core density contribution $\rho_\mathrm{core}^\mathrm{atom}(\vec{r})$ is then constructed by `unscreening' $v_l^\mathrm{total}(\vec{r})$ by subtracting the Hartree and XC contributions associated with the valence charge density $v_{Hxc}[\rho_\mathrm{val}](\vec{r})$ \cite{Hamann_Schulter_Chang:1979,Louie_NLCCs:1982}:
\begin{equation}
	v_l^\mathrm{ion}(\vec{r}) = v_l^\mathrm{total}(\vec{r}) -
	v_{H}[\rho_\mathrm{val}](\vec{r}) - \vxcs[\rho_\mathrm{val}](\vec{r}).
	\label{eq:uncorrected_nlcc}
\end{equation}
As pointed out first by Louie \textit{et al.}\ \cite{Louie_NLCCs:1982}, there is an implied linearization in this procedure:
\begin{equation}
	v_{Hxc}[\rho_\mathrm{total}](\vec{r}) = v_{Hxc}[\rho_\mathrm{core}](\vec{r}) + v_{Hxc}[\rho_\mathrm{val}](\vec{r}),
\end{equation}
where $v_{Hxc}[\rho](\vec{r}) = v_H[\rho](\vec{r}) + \vxcs[\rho](\vec{r})$.

Although this is true for the Hartree potential, the XC potential is typically nonlinear in the density and thus simply replacing the total charge density with the pseudovalence density in a solid state calculation can be only be an approximation.
In particular, the pseudopotential depends on the valence configuration $v_l^\mathrm{ion}(\vec{r})$ used to generate it.
This greatly hampers its transferability, especially when there is a significant overlap between
$\rho_\mathrm{core}(\vec{r})$ and $\rho_\mathrm{val}(\vec{r})$. Fortunately this can be corrected through the use of nonlinear core corrections (NLCCs), in which the following expression is used in place of Eq.\ \eqref{eq:uncorrected_nlcc}:
\begin{equation}
	v_l^\mathrm{ion}(\vec{r}) = v_l^\mathrm{total}(\vec{r}) -
	v_\text{H}[\rho_\mathrm{val}](\vec{r}) - \vxcs[\rho_\mathrm{val}+\rho_\mathrm{core}](\vec{r}),
\end{equation}
as discussed in Refs.\ \onlinecite{Louie_NLCCs:1982} and \onlinecite{Porezag_NLCCs:1999}.

\begin{figure*}[htbp!]
	\includegraphics[width=\linewidth]{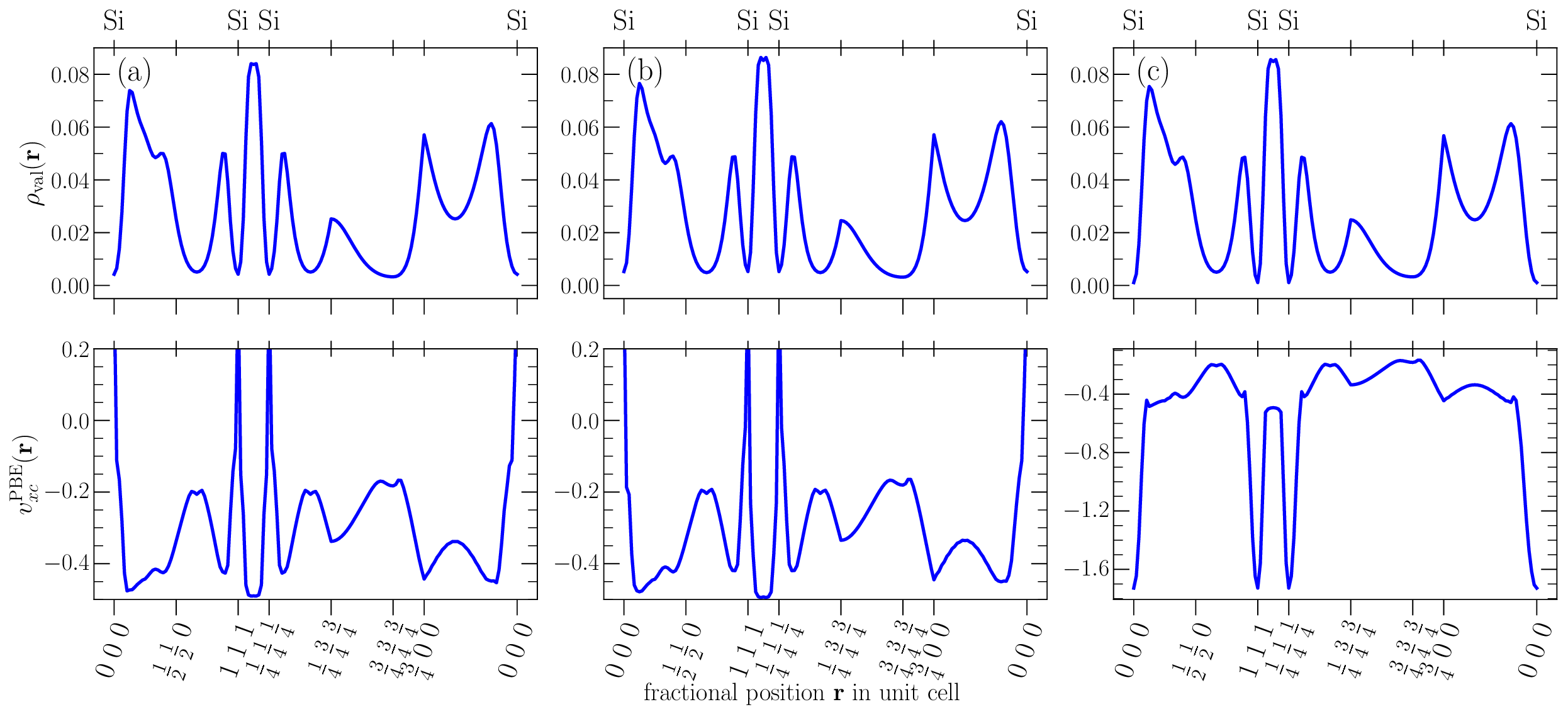}
	\caption{Plot of the (valence) charge density $\rho_\mathrm{val}(\vec{r})$ and the PBE XC potential $\vxcs^\mathrm{PBE}(\vec{r})$ in bulk Si for
		(a) TN pseudopotentials,
		(b) MHL pseudopotentials, neither of which include nonlinear core corrections (NLCCs),
		and
		(c) NCP19 on-the-fly-generated pseudopotentials in \textsc{castep} (see main text) with NLCCs.
		$\rho_\mathrm{val}(\vec{r})$ and $\vxcs^\mathrm{PBE}(\vec{r})$ are plotted along the path shown in Fig.\ \ref{fig:Si_locpot_compare} (a).
		The valence configuration for all three pseudopotentials is identical (see Table \ref{table:pspot_compare}).
	}
	\label{fig:nlcc_compare}
\end{figure*}

In our work, we find that the omission of NLCCs in the pseudopotential affects the shape of the XC potential $\vxcs(\vec{r})$, particularly near ions.
To illustrate the effects of NLCCs, we performed calculations for bulk Si with the PBE functional using three sets of pseudopotentials:
\begin{enumerate}
	\item the TN Dirac-Fock pseudopotentials used for all calculations in this work,
	\item the norm-conserving pseudopotentials of M.-H.\ Lee \cite{Lee_CASTEP_recpots_thesis:1995}, hereafter referred to as MHL \footnote{Distributed with \textsc{castep} as \texttt{Si\_00.recpot}},
	\item the OTFG pseudopotentials from the NCP19 library within \textsc{castep}.
\end{enumerate}
The TN and MHL pseudopotentials do not include NLCCs, while the NCP19 pseudopotentials do \cite{new_castep_pspots:2016}.
We further note that all these pseudopotentials use the same valence configuration as given in Table \ref{table:pspot_compare}.
The results are shown in Fig.\ \ref{fig:nlcc_compare}.
The valence charge densities $\rho_\mathrm{val}(\vec{r})$ of all three pseudopotentials are comparable, as expected since they use the same valence configuration in their generation.
However, the XC potentials for the TN and MHL pseudopotentials in Figs.\ \ref{fig:nlcc_compare}(a) and (b) exhibit spikes near the nuclei that make the $\vxcs^\mathrm{PBE}(\vec{r}) > 0$ (where in reality, $\vxcs^\mathrm{PBE}(\vec{r})\leq0$).
The inclusion of NLCCs results in similar spikes, although they are now negative, as shown in Fig.\ \ref{fig:nlcc_compare}. We note that away from ions, i.e.\ outside the core radius of the pseudopotential, $\vxcs^\mathrm{PBE}(\vec{r})$ is similar for the TN, MHL, and NCP19 pseudopotentials, since there is no contribution from $\rho_\mathrm{core}(\vec{r})$.

\bibliography{refs.bib}
\end{document}

\title{Supplemental Material: From Densities to Potentials: Benchmarking Local Exchange-Correlation Approximations}

\author{Visagan Ravindran}
\affiliation{Department of Physics, Durham University, South Road, Durham, DH1 3LE, United Kingdom}
\author{Clio Johnson}
\author{Neil D.\ Drummond}
\affiliation{Department of Physics, Lancaster University, Lancaster LA1 4YB, United Kingdom}
\author{Stewart J.\ Clark}
\author{Nikitas I.\ Gidopoulos}
\affiliation{Department of Physics, Durham University, South Road, Durham, DH1 3LE, United Kingdom}
\email[Corresponding author: ]{nikitas.gidopoulos@durham.ac.uk}

\date{30th October 2025}

\begin{abstract}
    In this supplementary material, we present the computational parameters used for the systems of study as well as some information on the preprocessing of densities prior to inversion. 
    We also demonstrate the convergence of the inversion algorithm as well as show the sum of the Hartree and exchange-correlation (XC) potentials $v_{Hxc}(\vec{r})$ in addition to the XC potential $\vxcs(\vec{r})$. 
    Finally, we also present Kohn-Sham band structures for select systems. 
\end{abstract}

\maketitle

\section{Computational parameters}
In Table~S\ref{suptable:comp_params}, we provide a list of the computational parameters for each material used for both the density functional theory (DFT) calculations in \textsc{CASTEP}\cite{Clark:05} code and quantum Monte Carlo (QMC) calculations using the \textsc{CASINO}\cite{Needs:20} code.
As stated in the main text, an initial Slater determinant for the Slater-Jastrow (SJ) trial wave function was obtained from a PBE\cite{PBE} calculation within \textsc{CASTEP} before the orbitals comprising the determinant were converted to a B-spline (blip) basis\cite{Hernandez_Gillan_blip:1997,Alfe_Gillan_blip:2004} for the subsequent QMC calculation. However, the plane-wave kinetic energy cutoff was used between both codes were the same as were the Trail-Needs (TN) pseudopotentials\cite{Trail:05}.

The size of the Monkhorst-Pack k-point grid used for this initial PBE calculation was commensurate with the size of the QMC computational supercell centered on the Baldereschi point. For the actual DFA calculations, the results of which are quoted in the main text, we used the k-point grids given in Table~S\ref{suptable:comp_params}.

\begin{table}[h]
	\begin{ruledtabular}
		\centering
		\begin{tabular}{lcdccc}
			Material&
			Structure&
			\multicolumn{1}{r}{\begin{tabular}[c]{@{}c@{}}Lattice \\Constant (\AA)\end{tabular}}&
			\begin{tabular}[c]{@{}c@{}}Plane-Wave\\Cutoff (Ha)\end{tabular}&
			\begin{tabular}[c]{@{}c@{}}DFA k-point\\Grid\end{tabular}&
			\begin{tabular}[c]{@{}c@{}}Pseudopotential valence\\ configuration\end{tabular}
			\\
			\hline
			Si& diamond& 5.43102& 50& $6\times6\times6$& Si - $3s^2 \; 3p^2$\\
			diamond& diamond& 3.56683 & 120& $6\times6\times6$& Si - $2s^2\; 2p^2$\\
			GaAs& zincblende& 5.65315 & 50& $6\times6\times6$&
			\begin{tabular}[c]{@{}c@{}}
				Ga - $4s^2\;4p^1$\\
				As - $4s^2\;4p^3$
			\end{tabular}\\
			Ge& diamond& 5.65791& 50& $6\times6\times6$& Ge-$4s^2\;4p^2$\\
			NaCl& rocksalt&5.64017 & 50& $6\times6\times6$&
			\begin{tabular}[c]{@{}c@{}}
				Na - $3s^1$
				\\
				Cl - $3s^2\;3p^5$
			\end{tabular} \\
			BaTiO\textsubscript{3}\footnote{Lattice parameter from Refs.\ \onlinecite{BaTiO3_Lat_Param_1} and \onlinecite{BaTiO3_Lat_Param_2}}&
			perovskite& 4.000& 220& $6\times6\times6$&
			\begin{tabular}[c]{@{}c@{}}
				Ba - $6s^2$
				\\
				Ti - $4s^2\;3d^2$
				\\
				O - $2s^2\;2p^4$
			\end{tabular} \\
			SrTiO\textsubscript{3}\footnote{Lattice parameters from Ref.\ \onlinecite{SrTiO3_Lat_Param}}&
			perovskite& 3.905& 220& $6\times6\times6$&
			Sr\footnote{Ti and O valence configuration same as BaTiO\textsubscript{3}.} - $5s^2$
			\\
			MnO\footnote{Lattice parameters from Ref.\ \onlinecite{TMO_lat_params}}&
			rocksalt& 4.45& 250& $5\times5\times3$& Mn\footnote{O valence configuration same as BaTiO\textsubscript{3}.} - $3d^5\;4s^2$
		\end{tabular}
	\end{ruledtabular}
	\label{suptable:comp_params}
	\caption{
		Calculation parameters used for the various materials studied in this work.
		The valence electronic configuration for the Trail-Needs (TN) pseudopotentials is also given.
		Experimental lattice parameters (quoted for conventional cells) are used from Ref.\ \cite{Madelung:1996} unless otherwise stated. The structures (in their conventional cell setting) correspond to the following space groups: diamond ($Fd\overline{3}m$), zincblende ($F\overline{4}3m$), rocksalt ($Fm\overline{3}m$), perovskite $(Pm\bar{3}m)$.
	}
\end{table}

\subsection{Baldereschi points}%
\label{section:baldereschi_mvp}

When calculating a QMC charge density, the supercell Bloch $k$-vector was chosen to lie at the Baldereschi mean value point (MVP) \cite{Baldereschi:73} of the supercell Brillouin zone in each case.  The MVP depends solely on the symmetry of the Bravais lattice of the simulation cell but has to be calculated numerically in general. Consider a smooth function $f(\vec{k})$ with the symmetry of the supercell reciprocal lattice.  We may write $f$ as a Fourier series,
\begin{equation}
    f\left(\vec{k}\right)=f_0+\sum_{n=1}^{\infty} f_n \sum_{\vec{R}\in\star_n} \exp\left(i\vec{R}\cdot\vec{k}\right) \equiv f_0 + \sum_{n=1}^{\infty} f_n S_n(\vec{k}),
\end{equation}
where $f_0$ is the mean of $f$, and $\star_n$ is the $n$th star of real supercell lattice points.  We seek a point $\vec{k}_\text{B}$ such that (i) $S(\vec{k}_\text{B})=0$ for $n=1$ (and $n=2$ and even $n=3$ if possible) and (ii) $|S_n(\vec{k}_\text{B})|$ is minimized for the first value of $n$ where it cannot be made zero.

We enforce condition (i) using the Newton-Raphson method to make $S_n(\vec{k}_\text{B})=0$ for $n=1$ to $Z$, where we try $Z=3$, then $Z=2$, then $Z=1$.
We then enforce condition (ii) by minimizing $K \sum_{n=1}^Z |S_n(\vec{k}_\text{B})|^2 + |S_{Z+1}(\vec{k}_\text{B})|^2$, where $K=10^8$ is a large constant.  Finally, we reimpose condition (i) using Newton-Raphson iteration again. The Broyden-Fletcher-Goldfarb-Shanno method is used to perform the minimization. A similar method of calculating MVPs has been documented in a recent paper by Stevanovic \cite{Stevanovic:24}.
The use of the Baldereschi point reduces momentum quantization effects for each supercell studied and hence facilitates extrapolation to infinite system size as detailed in Sec.\ III of the main text. For the insulators and semiconductors studied in this work, the use of supercell Baldereschi points provides a cheap alternative to averaging over twisted periodic boundary conditions on a supercell.

\section{Accumulation and preprocessing of densities}
We accumulate charge densities in reciprocal space in terms of their Fourier coefficients $\rho_{\vec{G}}$, where each primitive cell reciprocal lattice vector $\vec{G}$ lies within a cutoff radius $|2\vec{G}_\text{cut}|$ set by the energy $E_\text{cut}=|\vec{G}_{\text{cut}}|^2/2$. $E_\text{cut}$ here is the plane-wave cutoff energy used to define a basis set for the orbitals.
Our Fourier series convention is
\begin{equation}
    \rho\left(\vec{r}\right)=\frac{1}{\Omega}\sum\limits_{\vec{G}}\rho_{\vec{G}}\exp\left(-i\vec{G}\cdot\vec{r}\right),
\end{equation}
where $\Omega$ is the primitive cell volume and
\begin{equation}
    \rho_\vec{G} = \int\limits_{\Omega}\intd{\vec{r}} \, \rho\left(\vec{r}\right)\exp\left(i\vec{G}\cdot\vec{r}\right).
\end{equation}
The Fourier coefficients are normalized such that the $\vec{G} = \vec{0}$ coefficient is equal to the number of electrons in the primitive cell. This facilitates direct comparison between charge densities of different simulation cell sizes.
For spin-polarized systems such as MnO, we accumulate separate charge densities for spin-up and spin-down electrons.

Since \textsc{casino} and \textsc{castep} utilize slightly different formats of storing the charge density, some preprocessing is required to prepare the density for inversion. 
In particular, \textsc{CASTEP} stores the densities on real-space rectilinear grids.
To convert between the two formats, we used a fast Fourier transform (FFT) to calculate the real space density from the reciprocal space data produced by \textsc{CASINO}.

Note in some cases, due to Fourier aliasing, the inverse FFT from reciprocal space to real space resulted in a real-space density $\rho(\vec{r})$ that was strictly non-negative everywhere. Therefore, the Coulomb energy [see Eq.~(12) of the main text]
\begin{equation}
	U^{(n)}
	=
	\sum_\sigma
	\frac{1}{2}\iint\intd{r}\intd{r'}
	\frac{
		(\rho_\mathrm{QMC}(\vec{r})-\runden)
		[\rho_\mathrm{QMC}(\vec{r'})-\rho^{(n)}_\mathrm{QMC}(\vec{r})]
	}
	{|\vec{r}-\vec{r'}|}\geq 0,
\end{equation}
cannot be strictly minimized as the running density at the $n$th iteration,  $\runden \geq 0$, for all $\vec{r}$ while $\runden < 0$ for some $\vec{r}$.
The reason for the non-negativity of $\runden$ is due to the fact it is obtained from the running Kohn-Sham (KS) orbitals $\phi_i^{(n)}(\vec{r})$
\begin{equation}
	\runden = \sum_i^\mathrm{occ}|\phi_i^{(n)}(\vec{r})|^2,
\end{equation}
which are in turned obtained as the solution to the KS equations with the running potential $v_s^{(n)}(\vec{r})$
\begin{equation}
	\left(-\frac{1}{2}\nabla^2 + v_s^{(n)}(\vec{r})\right)\phi_i^{(n)}(\vec{r})
	=
	\varepsilon_i^{(n)}\phi_i^{(n)}(\vec{r})~.
\end{equation}
In these instances, an infimum $U$ is obtained, the true minimum naturally being $U=0$ when $\rho_\mathrm{QMC}(\vec{r}) = \runden$.  In practice, the negative density integrates to only a small fraction (around $10^{-5}$ to $10^{-4}$) of an electron.

\section{Convergence of inversion algorithm}
Here, we demonstrate the convergence of our inversion algorithm by inverting the QMC density in three systems: Si, BaTiO\textsubscript{3}, and MnO and comparing with the inversion of generalized Kohn-Sham (GKS) densities from non-local DFAs. The latter two materials are particularly instructive given that the density in these regions is very small, resulting in (likely spurious) oscillations in the inverted exchange-correlation (XC) potential $\vxcs(\vec{r})$, which would likely require many iterations of the algorithm to entirely remove.

\begin{figure}[h]
	\centering
	\includegraphics[width=\linewidth]{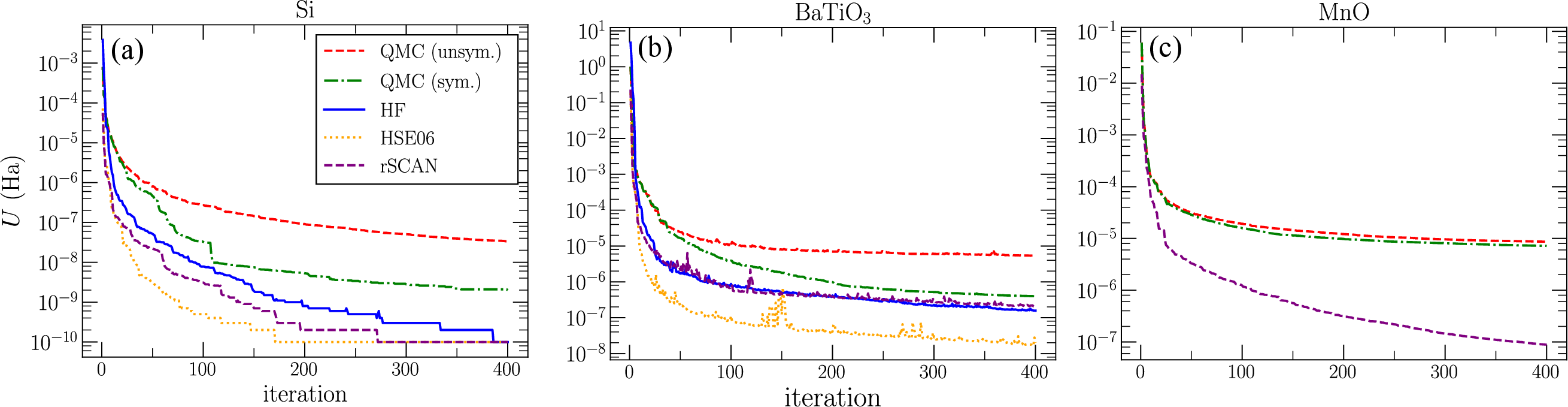}
	\caption{Convergence of the Coulomb energy $U^{(n)}$ (see main text) of bulk (a) Si, (b) BaTiO\textsubscript{3}, and (c) MnO for the inversion of various target densities.
	The green and red lines in each figure show the inversion explicitly symmetrized QMC density and the original unsymmetrized density respectively, as obtained from the $3\times3\times3$ computational supercell.
	}
	\label{supfig:U_convg}
\end{figure}

In Fig.~\ref{supfig:U_convg}, we monitor the convergence of the Coulomb energy $U^{(n)}$ for the inversion of the unsymmetrized and symmetrized QMC densities of bulk Si, BaTiO\textsubscript{3} in $3\times 3\times 3$ supercells.
Note that for MnO, we encountered the aforementioned numerical issues outlined in the previous section when performing the FFT, with the negative density integrating to $2\times10^{-4}$ electrons. In this instance, the minimum of $U=0$ cannot be reached, even in principle.

\begin{figure}[!h]
	\centering
	\includegraphics[width=\linewidth]{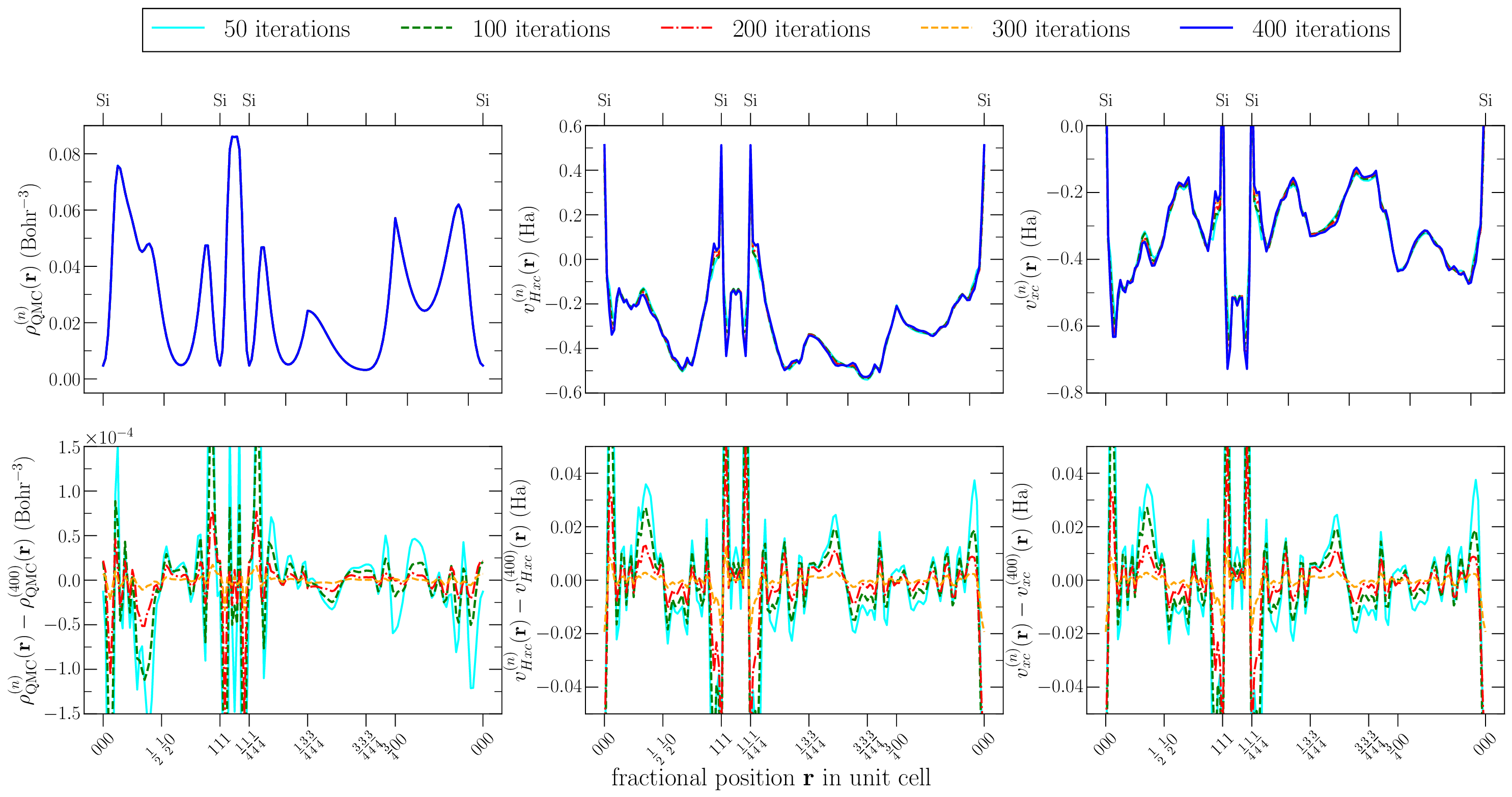}
	\caption{
		Convergence study of the inversion of the unsymmetrized QMC density in bulk Si using a $3\times 3\times 3$ supercell.
		The top panels give the running density $\runden$ as well as  $\runhxc=\runh+\runxc$, and XC, $\runxc$, potentials at the $n$th iteration, while the bottom panels give the difference in the respective quantities between the $i$th iteration and 400 iterations.
		The path through the unit cell is the same as that of Fig.~5 in the main text.}
	\label{supfig:Si_unsym_convg}
\end{figure}

\begin{figure}[!]
	\centering
	\includegraphics[width=\linewidth]{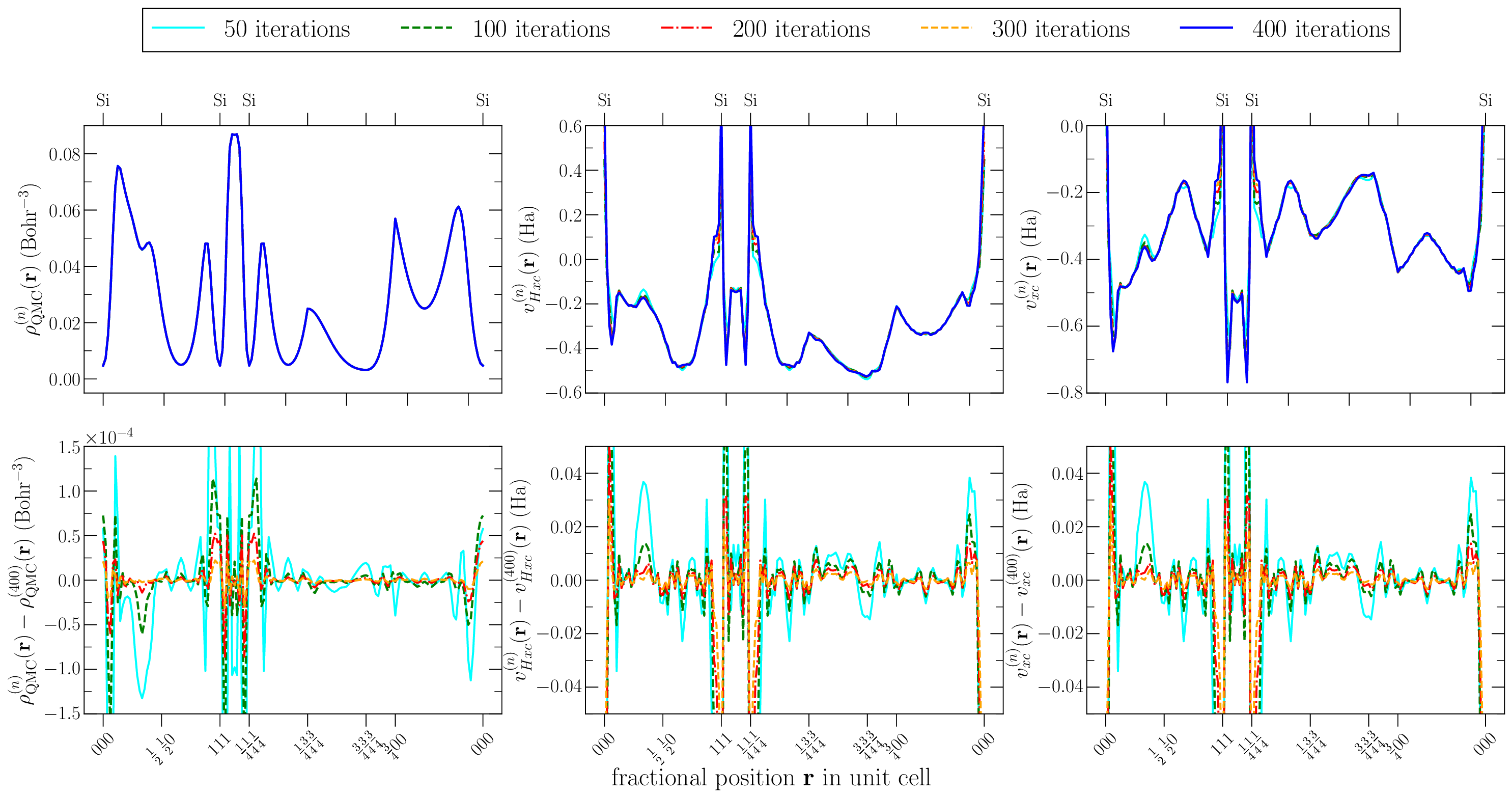}
	\caption{
		Same as Fig.~\ref{supfig:Si_unsym_convg} but for symmetrized density of bulk Si.}
	\label{fig:Si_sym_convg}
\end{figure}

A comparison of $\runden$ well as the Hartree+XC, $\runhxc=\runh+\runxc$, and XC, $\runxc$, potentials at the $n$th iteration is shown in Figs.~\ref{supfig:Si_unsym_convg} and \ref{fig:Si_sym_convg} for the unsymmetrized and symmetrized densities respectively for bulk Si where one can see that as the number of iterations is increased, both the $\runden$ and $\runhxc$, and by extension, $\runxc$ converge as shown by the small difference in the differences between these quantities and the quantities at 400 iterations in the bottom panel.
The faster convergence in the symmetrized case in Fig.~\ref{supfig:U_convg} (a) can be attributed to the fact that symmetrization procedure essentially removes some random error (noise) present in the density. The QMC density will satisfy the appropriate crystallographic symmetries if the runtime is sufficiently long, such that the difference between the raw QMC unsymmetrized density and symmetrized density would be small.

\begin{figure}[!]
	\centering
	\includegraphics[width=\linewidth]{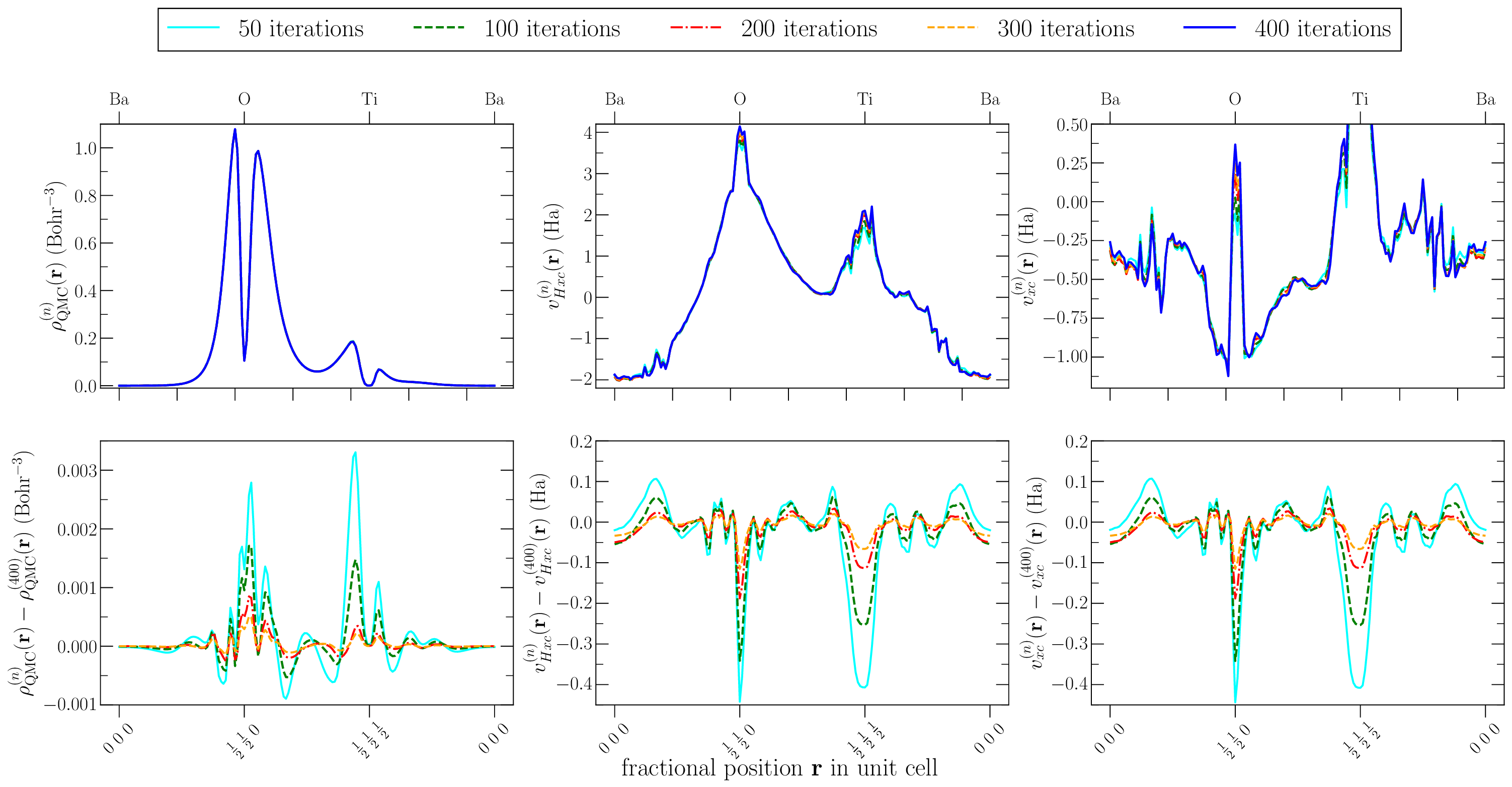}
	\caption{
		Convergence study of the inversion of the unsymmetrized QMC density in bulk BaTiO\textsubscript{3} using a $3\times 3\times 3$ supercell.
		The top panels give the running density $\runden$ as well as  $\runhxc=\runh+\runxc$, and XC, $\runxc$, potentials at the $n$th iteration, while the bottom panels give the difference in the respective quantities between the $i$th iteration and 400 iterations.
		The path through the unit cell is the same as that of Fig.~8 in the main text.}
	\label{supfig:BTO_unsym_convg}
\end{figure}

\begin{figure}[!]
	\centering
	\includegraphics[width=\linewidth]{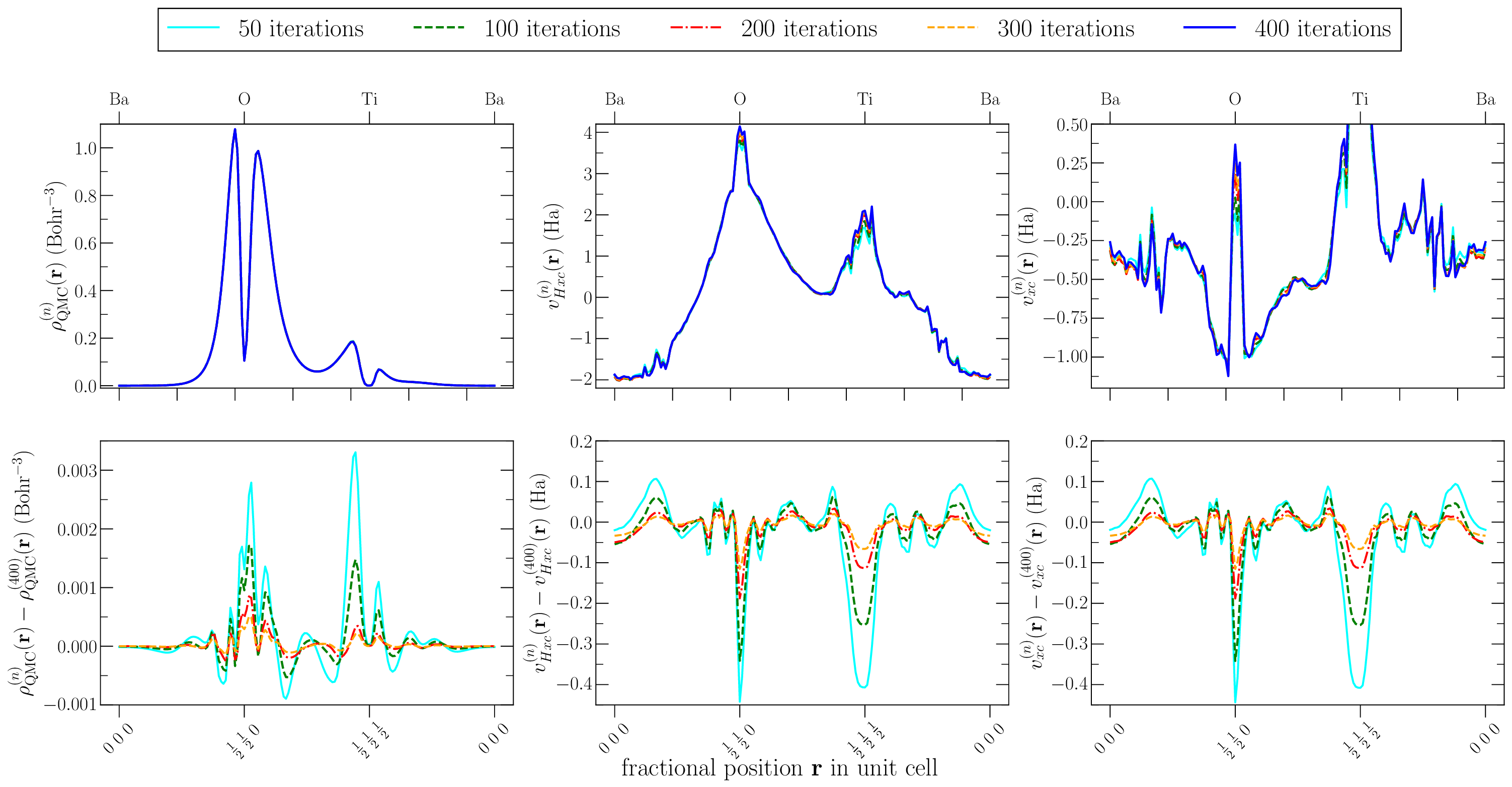}
	\caption{
		Same as Fig.~\ref{supfig:BTO_unsym_convg} but for symmetrized density of bulk BaTiO\textsubscript{3}.
	}
	\label{supfig:BTO_sym_convg}
\end{figure}

\begin{figure}[!]
	\centering
	\includegraphics[width=\linewidth]{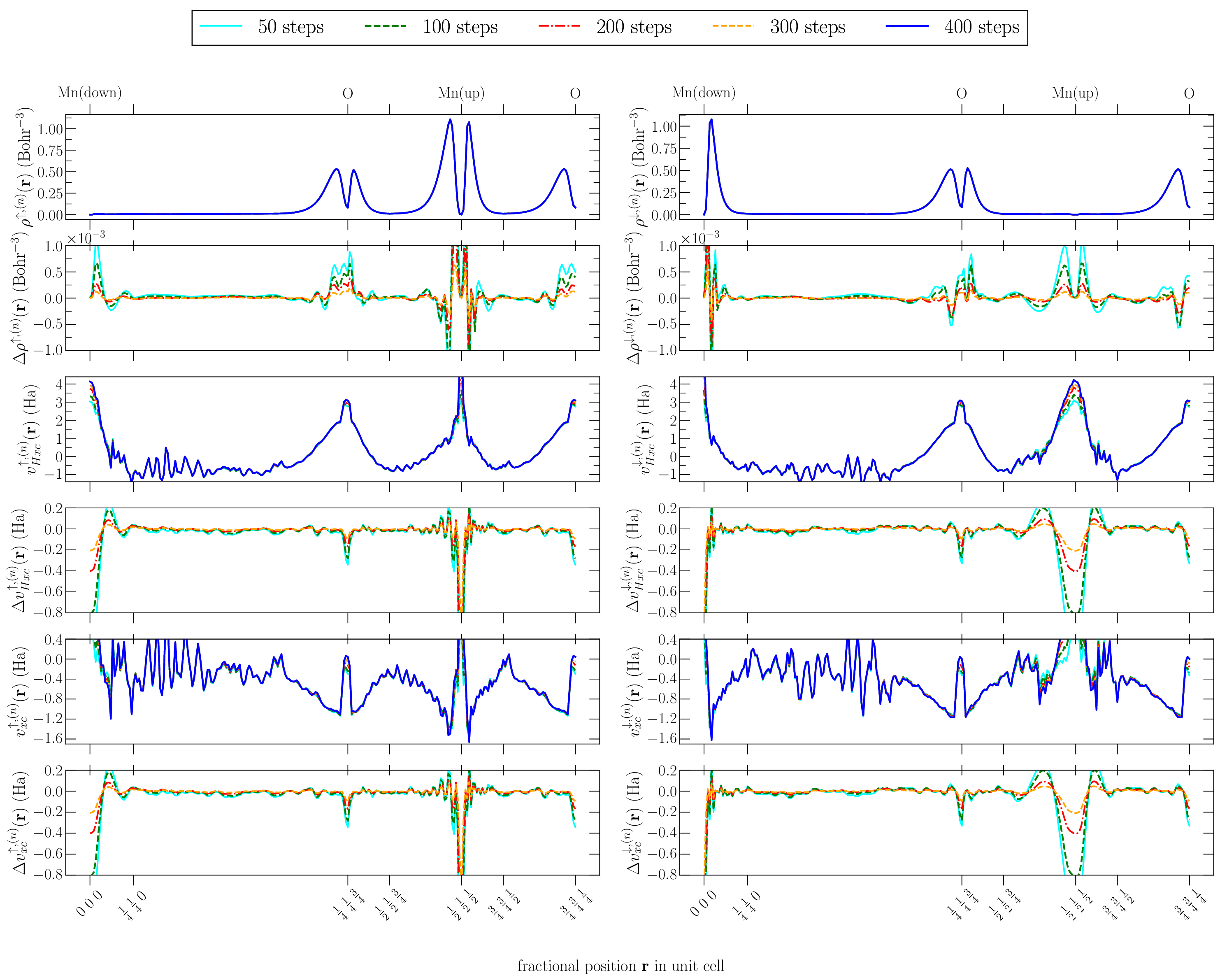}
	\caption{
		Convergence study of the inversion of the unsymmetrized QMC density in bulk MnO using a $3\times 3\times 3$ supercell.
		The density for each spin channel $\sigma$, $\rho^{\sigma,(n)}(\vec{r})$ at the $n$th iteration along with the sum of Hartree $v_H^{\sigma,(n)}(\vec{r})$ and XC potentials $v_{xc}^{\sigma,(n)}(\vec{r})$,
		$v_{Hxc}^{\sigma,(n)}(\vec{r})=v_H^{\sigma,(n)}(\vec{r})+v_{xc}^{\sigma,(n)}(\vec{r})$,
		are plotted along the same path through the unit cell as Fig.~9 in the main text.
		The difference between the $n$th iteration and 400 iterations is shown below the respective quantity, for instance \\
		$\Delta \rho^{\sigma,(n)}(\vec{r}) = \rho^{\sigma,(n)}(\vec{r}) - \rho^{\sigma,(400)}(\vec{r})$.
		The relevant spin-up and spin-down quantities are shown on the left and right panels respectively.
	}
	\label{supfig:MnO_unsym_convg}
\end{figure}

\begin{figure}[!]
	\centering
	\includegraphics[width=\linewidth]{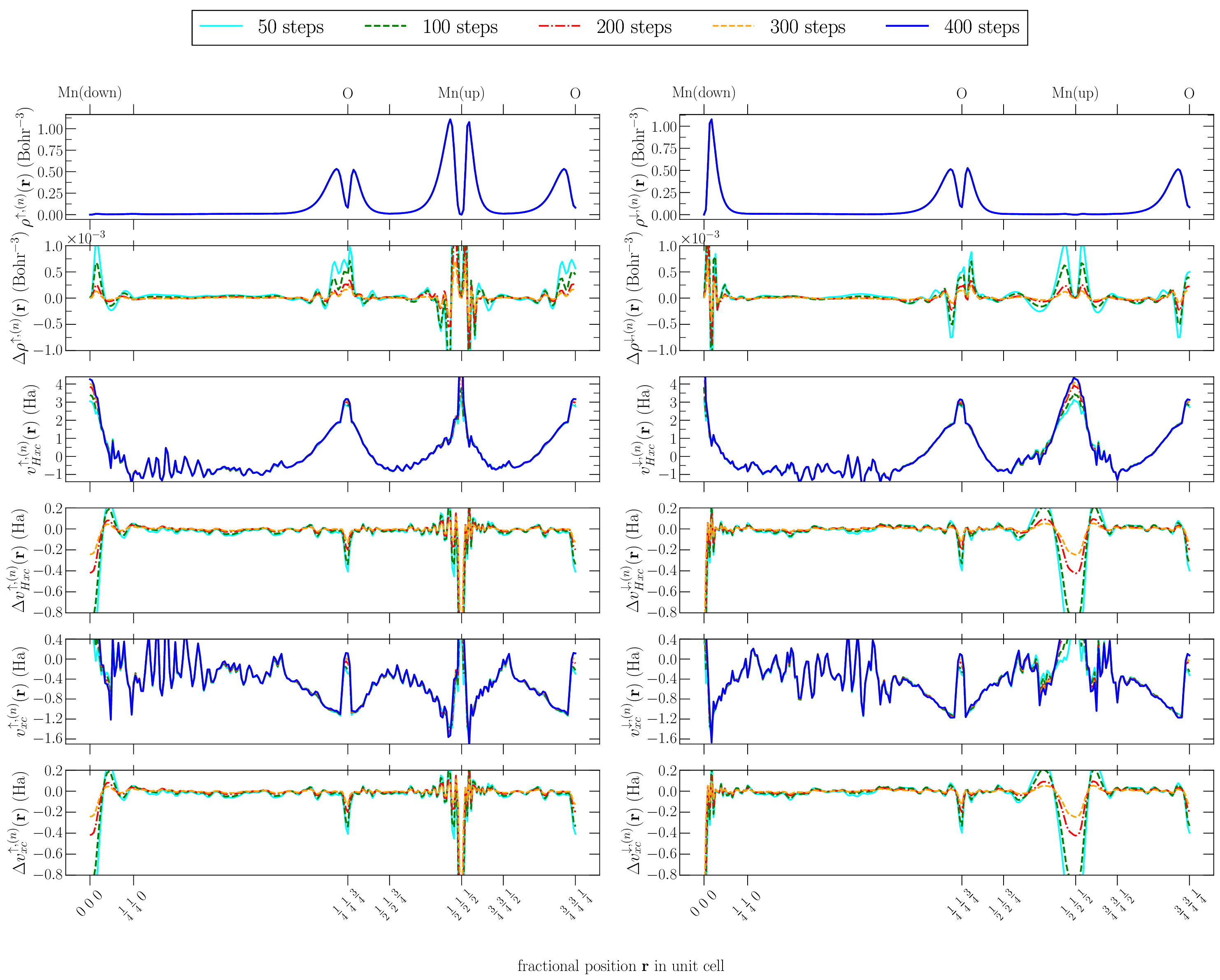}
	\caption{
		Same as Fig.~\ref{supfig:MnO_unsym_convg} but for the symmetrized QMC density in bulk MnO in a $3\times 3\times 3$ supercell.
	}
	\label{supfig:MnO_sym_convg}
\end{figure}

We show similar comparisons for bulk BaTiO\textsubscript{3} in Figs.~\ref{supfig:BTO_unsym_convg} and \ref{supfig:BTO_sym_convg} and MnO in Figs.~\ref{supfig:MnO_unsym_convg} and \ref{supfig:MnO_sym_convg} for the unsymmetrized and symmetrized cases respectively. Note the comparable convergence of both $v_{Hxc}(\vec{r})$ and $v_{xc}(\vec{r})$.

Finally, we mention that the actual observables within the KS system such as the KS eigenvalues appear to converge faster than potential and in particular, the oscillations observed in the potential, particularly in the low density region appear to have a smaller effect on the eigenvalues and especially the Kohn-Sham gap.
Table~\ref{suptable:KS_gap_convergence} gives the value of the KS indirect gap for bulk Si, BaTiO\textsubscript{3}, and MnO using the same methods to generate the target density as in Fig.~\ref{supfig:U_convg} after a set number of iterations has been completed. We note that the gap is converged to within a few meV within a 100 iteration of the inversion algorithm although the $v_{Hxc}(\vec{r})$ and $v_{xc}(\vec{r})$ are not, notably in MnO (see Fig.~\ref{supfig:MnO_sym_convg}).
\begin{table*}[b]
	\begin{ruledtabular}
		\begin{tabular}{c|x{5}x{5}x{5}x{5}x{5}|x{5}x{5}x{5}x{5}x{5}|x{5}x{5}x{5}}
			{}&
			\multicolumn{5}{c|}{Si}&
			\multicolumn{5}{c|}{BaTiO3\textsubscript{3}}&
			\multicolumn{3}{c}{MnO}
			\\
			iterations&
			\multicolumn{1}{c}{rSCAN}& \multicolumn{1}{c}{HSE06}&
			\multicolumn{1}{c}{HF}&
			\multicolumn{1}{c}{\makecell{QMC\\(unsym.)}}&
			\multicolumn{1}{c|}{\makecell{QMC\\(sym.)}}&
			\multicolumn{1}{c}{rSCAN}& \multicolumn{1}{c}{HSE06}&
			\multicolumn{1}{c}{HF}&
			\multicolumn{1}{c}{\makecell{QMC\\(unsym.)}}&
			\multicolumn{1}{c|}{\makecell{QMC\\(sym.)}}&
			\multicolumn{1}{c}{rSCAN}&
			\multicolumn{1}{c}{\makecell{QMC\\(unsym.)}}&
			\multicolumn{1}{c}{\makecell{QMC\\(sym.)}}
			\\
			\hline
			50&
			0.69943&        0.70102&        1.19952&        0.78217&        0.79644&        2.57161&        2.77746&        4.78474&        3.22446&        3.25438&        3.24082&        2.95549& 2.96100\\
			100&     0.69913&        0.70111&        1.20005&        0.78371&        0.79954&        2.55656&        2.77418&        4.76833&        3.21018&        3.24110&        3.24290&        2.94780& 2.95536\\
			200&     0.69890&        0.70126&        1.20064&        0.78450&        0.80043&        2.55153&        2.77470&        4.76743&        3.21212&        3.23981&        3.24437&        2.94228& 2.95001\\
			300&     0.69891&        0.70135&        1.20069&        0.78492&        0.80074&        2.55283&        2.77430&        4.76678&        3.20625&        3.24098&        3.24471&        2.94018& 2.94919\\
			400&     0.69888&        0.70139&        1.20068&        0.78504&        0.80094&        2.55362&        2.77500&        4.76768&        3.20387&        3.24189&        3.24465&        2.93947& 2.94861\\
		\end{tabular}
	\end{ruledtabular}
	\label{suptable:KS_gap_convergence}
	\caption{
		Calculated KS gaps (in eV) obtained after a set number of iterations in the inversion algorithm for the respective target density (cf.\ Table~\ref{supfig:U_convg}).
		For the QMC gaps, the densities are obtained using $3\times3\times3$ computational supercells and we distinguish between the gap obtained using the explicitly symmetrized (sym.)\ and unsymmetrized (unsym.)\ densities, i.e.\ raw data.
	}
\end{table*}

\clearpage
\section{Contribution of Hartree potential to the Kohn-Sham potential}
In the main text as well as previous work\cite{den_inv_Wasserman:2021,den_inv_Aouina_Reining_QMC:2023}, we note that different DFAs can yield different densities even but comparatively larger differences in the KS potential $v_s(\vec{r})$. The similarity of densities however leads to very similar Hartree potentials $v_H(\vec{r})$ between different DFAs. Consequently, the difference between the KS potentials $v_s(\vec{r})^\mathrm{DFA}$ of different DFAs emerges primarily from the differences in $v_{xc}(\vec{r})$.

\begin{figure}[h]
		\centering
		\includegraphics[width=\linewidth]{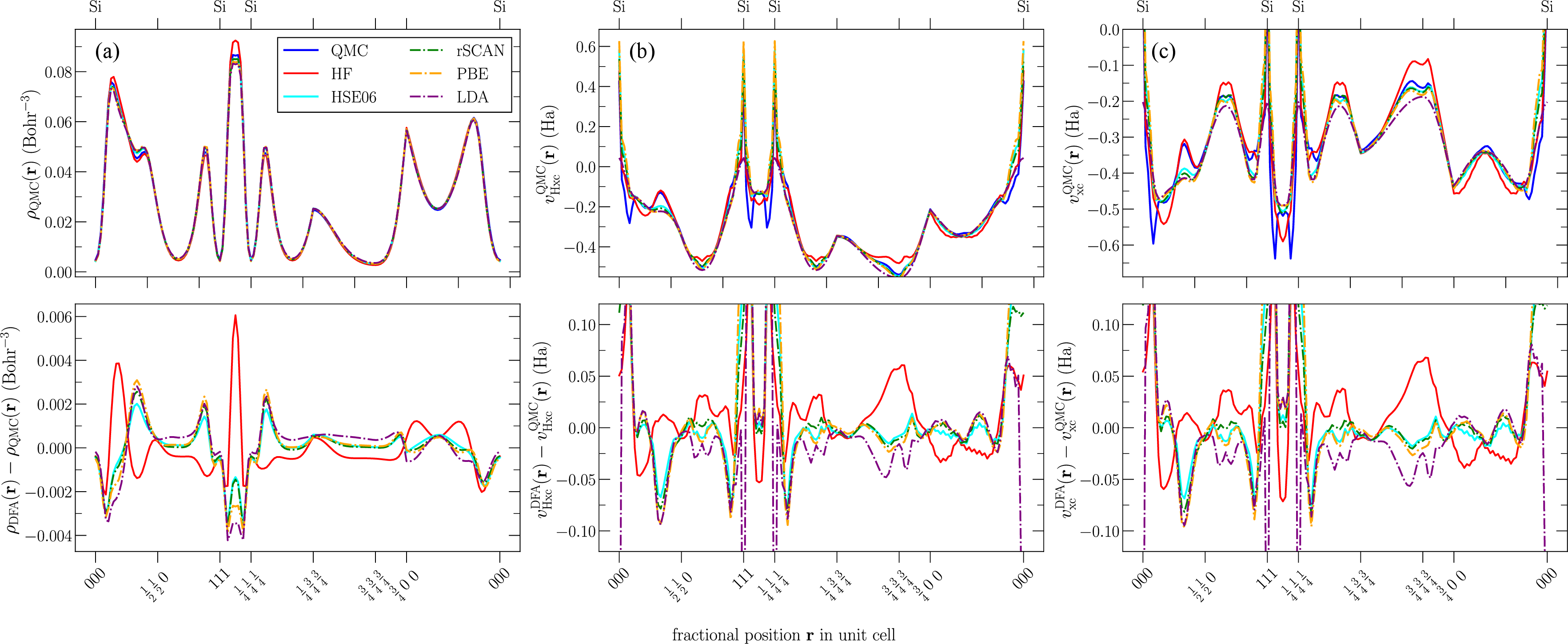}
		\caption{
			Top: (a) Density, (b) sum of Hartree and XC potential $v_{Hxc}(\vec{r})=v_H(\vec{r})  + \vxcs(\vec{r})$, and (c) XC potential $\vxcs(\vec{r})$ for various methods in bulk Si plotted along the same path through the unit cell as Fig.~5.
			The bottom panels give the difference between DFAs and the QMC result for each respective quantity.
		}
		\label{supfig:si_vhxc}
\end{figure}

\begin{figure}[h]
	\centering
	\includegraphics[width=\linewidth]{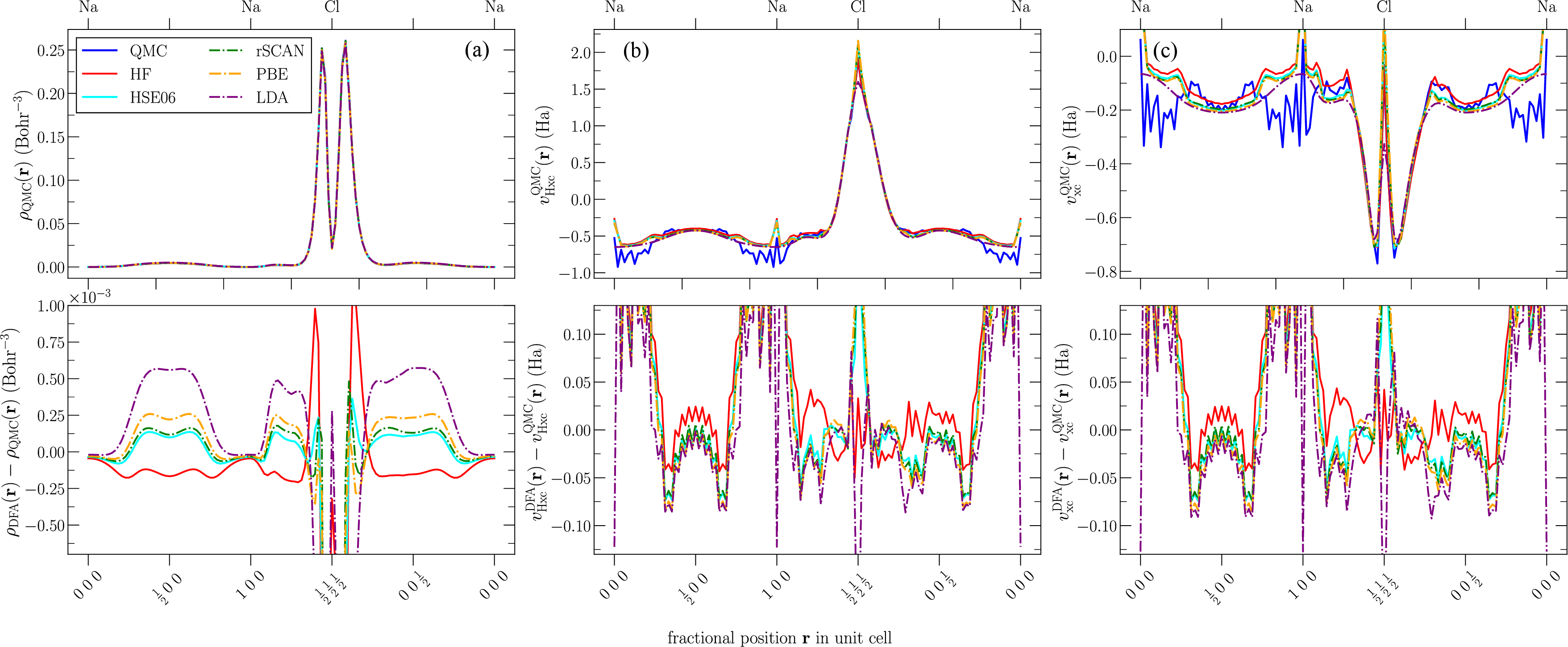}
	\caption{
		Same as Fig.~\ref{supfig:si_vhxc} but for bulk NaCl.
	}
	\label{supfig:nacl_vhxc}
\end{figure}

This is shown in Figs.~\ref{supfig:si_vhxc} and \ref{supfig:nacl_vhxc} for bulk Si and NaCl.
One can see that the differences between $\vxcs(\vec{r})$ for the various DFAs is comparable to the difference in $v_{Hxc}(\vec{r})$ .
Separately one can see this behavior is similar to the convergence of both $\vxcs(\vec{r})$ and $v_{Hxc}(\vec{r})$ in the inversion algorithm as discussed in the previous section.

\section{Comparison of Kohn-Sham band structures}
Here, we provide both KS and GKS band structures for a few select systems.
As discussed in Ref.~\onlinecite{Ravindran:2024}, the dispersion between GKS and KS band structures for a given DFA is similar with the difference primarily in the band gap itself.
In the case of KS, one needs to incorporate the XC derivative discontinuity correction in order to get the correct fundamental gap\cite{PPLB_Delta_XC:1982}
\begin{equation}
	E_\mathrm{g} = E_{\mathrm{g},s} + \Delta_{xc},
\end{equation}
This is not the case for the GKS where the $\Delta_{xc}$ correction is included within the GKS gap\cite{GKS_OEP_SCAN_Perdew_2016,band_gaps_in_GKS_2017}.
Consequently, the band gaps obtained from the GKS band structure are larger than in KS, with the difference between a GKS and a KS treatment depending on the strength of the nonlocality of the DFA (with an entirely local/semilocal DFA in the density yielding identical results).

\begin{figure}[t]
	\centering
	\includegraphics[width=\linewidth]{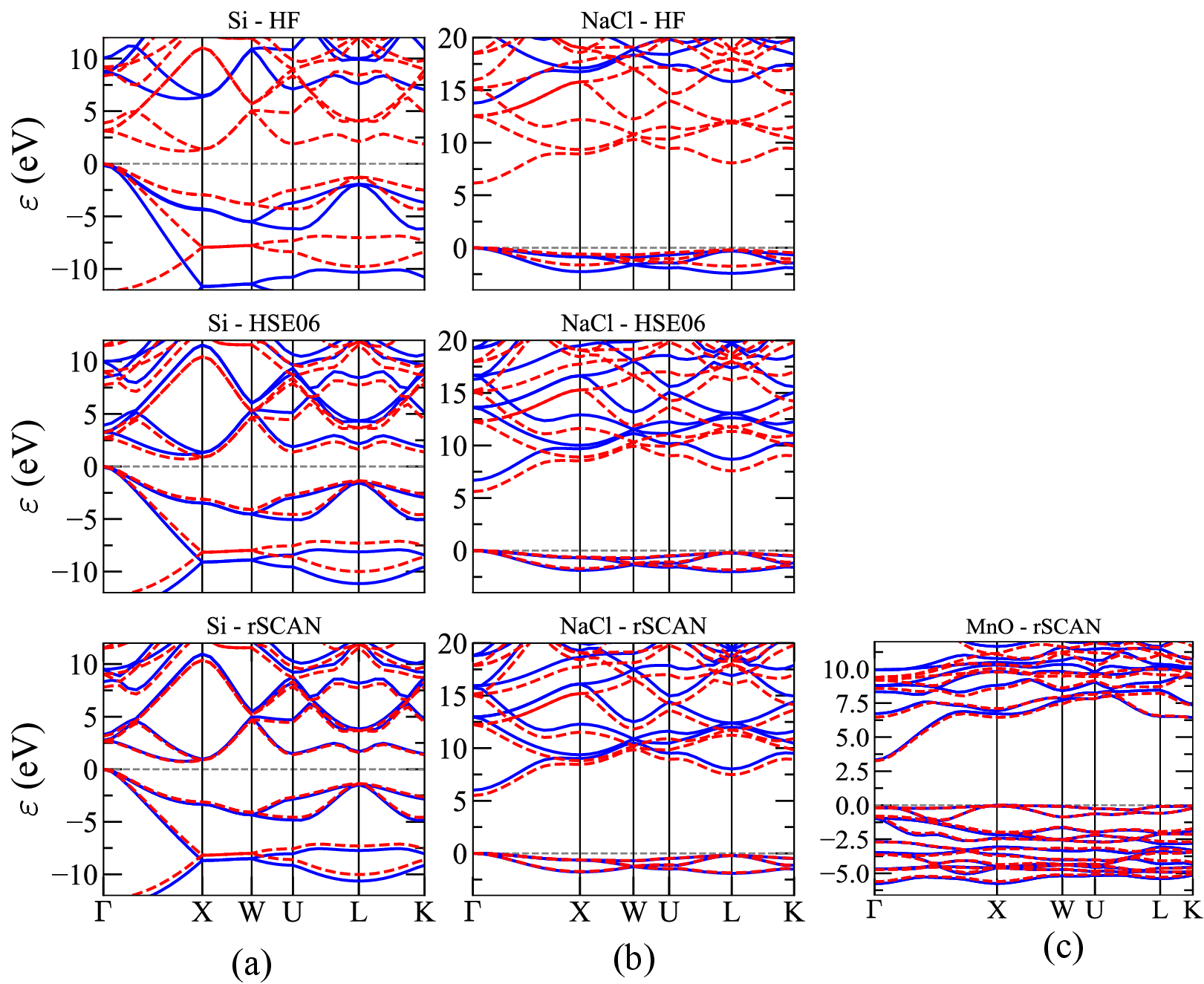}
	\caption{
		Generalized Kohn-Sham (GKS) (in solid blue) and Kohn-Sham (KS) (in dotted red) band structures of (a) Si and NaCl and (b) MnO\@.
		Note due to the high plane-wave cutoff, the HF and HSE06 calculations were too costly to run.
		The energy scale in all band structures has been set such that the Fermi energy of the KS band structure is at 0 eV while the GKS band structure has been shifted such that the KS and GKS valence band maxima coincide.
	}
	\label{supfig:gks_vs_ks}
\end{figure}
This can be seen for instance in Figs.~\ref{supfig:gks_vs_ks} (a) and (b) where we compare the KS and GKS band structures of HF, HSE06, and rSCAN for Si, NaCl, and MnO, where the KS band structure is obtained via inversion of the GKS target density.
GKS-HF and GKS-HSE06 for MnO proved too costly to run from a memory standpoint due to the high plane-wave cutoff required to converge the basis set and therefore we only present results for MnO for the rSCAN functional.
Note in the case of KS-HF, otherwise known as local Fock exchange (LFX) in previous work\cite{Hollins:2017_LFX,Hollins:2017_LFX_metals,Ravindran:2024}, there is only an exchange discontinuity $\Delta_{x}$ due to the lack of correlation.

\begin{figure}[!h]
	\centering
	\includegraphics[width=\linewidth]{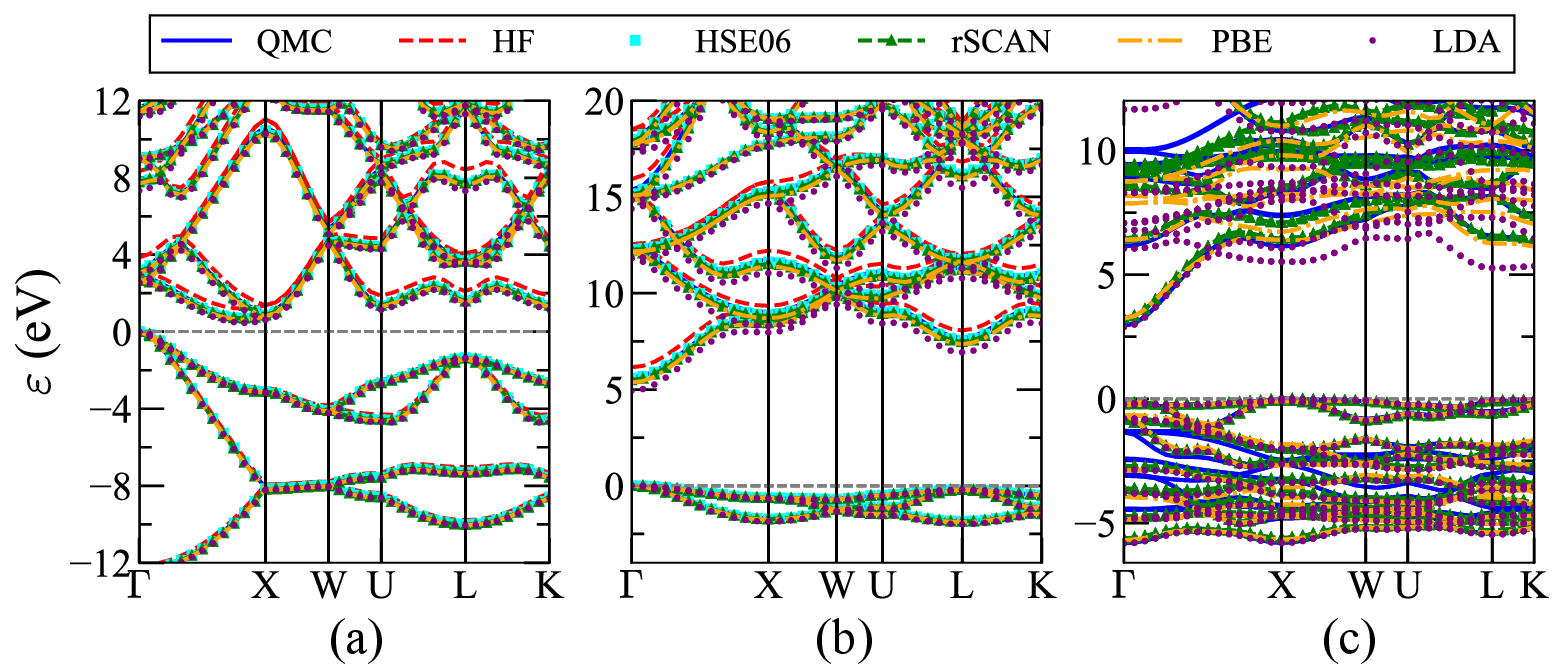}
	\caption{Kohn-Sham band structures for (a) Si, (b) NaCl, and (c) MnO using various methods. The energy scale has been such that the valence band maxima of each method coincide.}
	\label{supfig:ks_bandstrucs}
\end{figure}

\begin{figure}[!h]
	\centering
	\includegraphics[width=\linewidth]{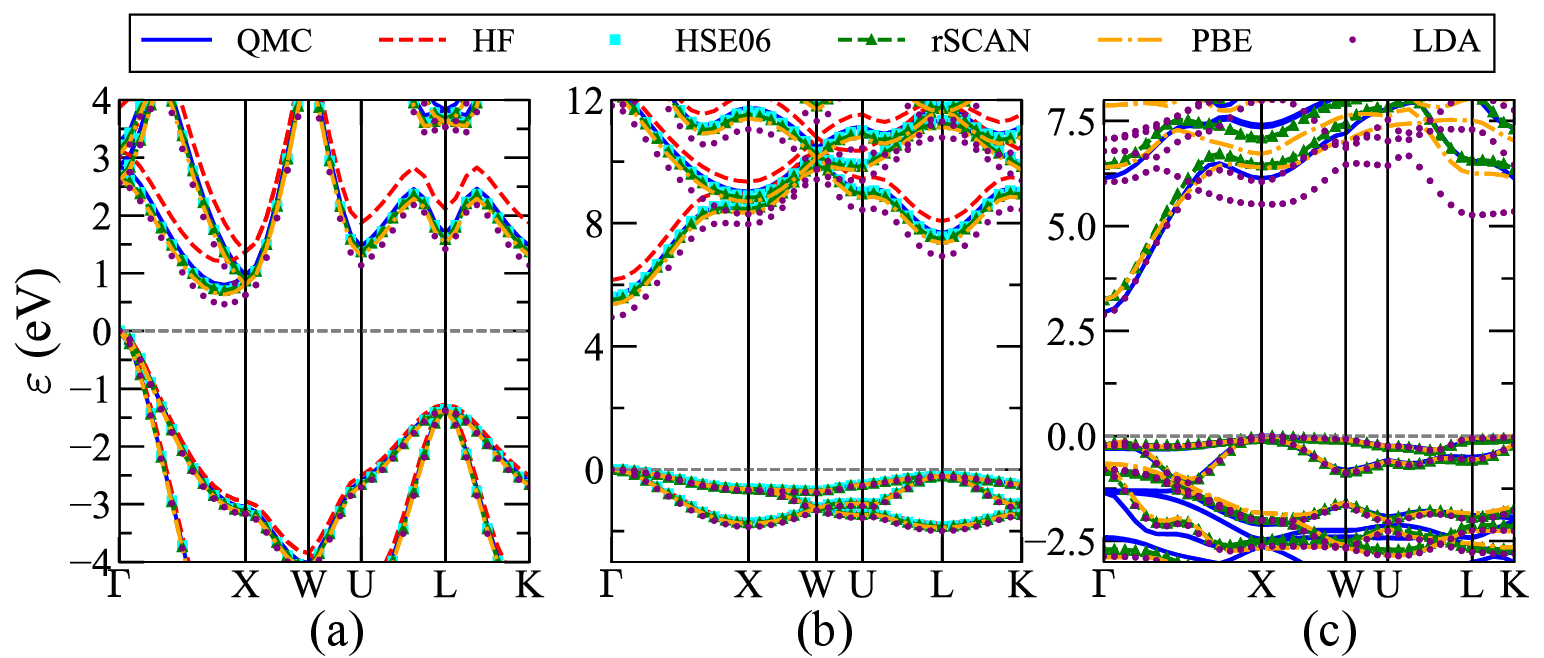}
	\caption{Zoomed-in version of Fig.~\ref{supfig:ks_bandstrucs}}
	\label{supfig:ks_bandstrucs_zoomed}
\end{figure}

Finally, for completeness, we also compare the KS band structure for each method for these systems in Fig.~\ref{supfig:ks_bandstrucs}.
A zoomed in version of Fig.~\ref{supfig:ks_bandstrucs} is given in Fig.~\ref{supfig:ks_bandstrucs_zoomed}.
In the case of Si and NaCl, the KS gaps for PBE, rSCAN, and HSE06 are comparable to each other and the KS-QMC gap although the KS-QMC gap is slightly larger.
On the other hand as pointed out in the main text, MnO had a larger KS-rSCAN gap compared to the KS-QMC gap although we stress that the GKS gap was still larger as expected. Furthermore, we point out that while the overall dispersion of bands remains similar across DFAs for Si and NaCl, greater differences can be seen for MnO\@.

\bibliography{refs.bib}